\documentclass[reprint,aps,pra,twocolumn,superscriptaddress,
longbibliography,floatfix,secnumarabic,footinbib]{revtex4-1}%
\usepackage{graphicx}
\usepackage{amsfonts}
\usepackage{amsmath}
\usepackage{ltxtable}
\usepackage{CJK}
\usepackage{mdwmath}
\usepackage{color}
\usepackage{appendix}
\usepackage{mathtools}
\usepackage{amssymb}
\usepackage{bm}
\usepackage{bbold}
\usepackage{booktabs}
\usepackage{threeparttable}
\usepackage[normalem]{ulem}
\usepackage{makecell}
\usepackage[colorlinks,allcolors=blue]{hyperref}

\begin{document}

\title{Deterministic All-versus-nothing Proofs of Bell Nonlocality Induced from Qudit Non-stabilizer States}

\author{Wenjing Du}
\affiliation{School of Physics and Information Technology, Shaanxi Normal University, Xi'an 710119, China}
\author{Di Zhou}
\affiliation{School of Physics and Information Technology, Shaanxi Normal University, Xi'an 710119, China}
\author{Kanyuan Han}
\affiliation{School of Mathematics and  Statistics, Shaanxi Normal University, Xi'an 710119, China}
\author{Hui Sun}
\affiliation{School of Physics and Information Technology, Shaanxi Normal University, Xi'an 710119, China}
\author{Huaixin Cao}
\affiliation{School of Mathematics and  Statistics, Shaanxi Normal University, Xi'an 710119, China}
\author{Weidong Tang}
\email{wdtang@snnu.edu.cn}
\affiliation{School of Mathematics and  Statistics, Shaanxi Normal University, Xi'an 710119, China}

\begin{abstract}		

Recently, a kind of deterministic all-versus-nothing proof of Bell nonlocality induced from the qubit non-stabilizer state was proposed, breaking the tradition that deterministic all-versus-nothing proofs are always derived from stabilizer states. A trivial generalization to the qudit ($d$ is even) version is by using a special basis map, but such a proof can still be reduced to the qubit version.
So far, whether high dimensional non-stabilizer states can induce nontrivial deterministic all-versus-nothing proofs of Bell nonlocality remains unknown. Here we present an example induced from a specific four-qudit non-stabilizer state (with $d=4$), showing that such proofs can be constructed in high dimensional scenarios as well.

\end{abstract}

\maketitle

\section{Introduction}

Bell nonlocality shows us a counterintuitive feature that correlations shared by some space-like entangled quantum systems cannot be simulated by any local hidden variable (LHV) theory\cite{Bell,Bell2,CHSH,Bell-Nonlocality-RMP-2014}. So far, there are mainly two ways to demonstrate Bell nonlocality: one is by Bell inequalities, and the other is by logical contradictions (sometimes also referred to as the demonstration of Bell locality without inequalities). Some proofs of Bell locality  without inequalities can rule out LHV theories with a success probability of $100\%$ (can even demonstrate Bell nonlocality in a single run of the experiment), and can be considered as deterministic all-versus-nothing (DAVN) proofs of Bell nonlocality\cite{all-vs-nothing}. The GHZ paradox\cite{GHZ1989,GHZ1990} and Cabello's proof\cite{Cabello2001} based on two Bell states are typical examples thereof. There are also some proofs (without inequalities) which can provide probabilistic all-versus-nothing  proofs of Bell's theorem, i.e., they can show Bell nonlocality with a success probability less than $100\%$, such as Hardy's paradox\cite{Hardy92,Hardy93}.

It is known the states capable of inducing DAVN  proofs of Bell nonlocality are demanding. In many GHZ paradoxes and Cabello's proof, the involved states are stabilizer states. Recently, one of authors of this work  found that some qubit non-stabilizer states can induce DAVN proofs of Bell nonlocality\cite{Tang-DAVN-2022} as well. Besides,  by using some basis map (e.g. $|\uparrow\rangle\rightarrow|0\rangle$, $|\downarrow\rangle\rightarrow|d/2-1\rangle$, where $d$ is even, $|\uparrow\rangle$ and $|\downarrow\rangle$ are two eigenstates of normal Pauli-$Z$ operator $\sigma_z$ with eigenvalues $+1$ and $-1$ respectively), one can naturally get some qudit versions of those proofs based on the qubit examples.
But these proofs can only be considered as a kind of dimensional reducible DAVN proofs rather than genuine $d$-dimensional ones, and thus we call them ``trivial qudit DAVN proofs" from non-stabilizer states.
So far, whether qudit ($d>2$) non-stabilizer states can induce non-trivial (genuine $d$-dimensional) DAVN proofs remains unknown.  To address this problem requires sophisticated calculations.
In this work, we shall report the first non-trivial DAVN proof  of Bell  nonlocality based on a qudit ($d=4$) non-stabilizer state, giving a positive answer to the problem.

\section{Main results}

First, let us consider a trivial generalization of a DAVN proof  of Bell  nonlocality from a qubit non-stabilizer state. Ref.\cite{Tang-DAVN-2022}  shows us that
the non-stabilizer state
\begin{align}\label{qubit-PCG-state}
    |\Psi_4\rangle=\frac{1}{\sqrt{7}}(|\uparrow\uparrow\uparrow\uparrow\rangle-|S(2,2)\rangle)
\end{align}
can induce a DAVN proof, where $|S(2,2)\rangle\equiv|\uparrow\uparrow\downarrow\downarrow\rangle+|\uparrow\downarrow\uparrow\downarrow\rangle
+|\uparrow\downarrow\downarrow\uparrow\rangle+|\downarrow\uparrow\uparrow\downarrow\rangle+|\downarrow\uparrow\downarrow\uparrow\rangle
+|\downarrow\downarrow\uparrow\uparrow\rangle$ is an un-normalized symmetric state.

Denote by $Z=\sum_{k=0}^{d-1}e^{i\frac{2k\pi}{d}}|k\rangle\langle k|$ and $X=\sum_{k=0}^{d-1}|k\oplus1\rangle\langle k|$
two qudit  Pauli matrices, where $\oplus$ stands for addition modulo $d$. Let $|0\rangle,|1\rangle,|2\rangle,\cdots,|d-1\rangle$ be the eigenstates of $Z$ with eigenvalues  $1, e^{i\frac{2\pi}{d}}, e^{i\frac{4\pi}{d}},\cdots, e^{i\frac{2(d-1)\pi}{d}}$ respectively.
One can also check that $X$ has the same eigenspectrum, and the eigenstate corresponding to the eigenvalue $e^{i\frac{2(m-1)\pi}{d}}$ can be
described as $|\theta_m\rangle=\frac{1}{\sqrt{d}}\sum_{k=0}^{d-1}e^{i\frac{2mk\pi}{d}}|k\rangle$, where $m\in\{0,1,\cdots,d-1\}$.

By replacing $|\uparrow\rangle$ and $|\downarrow\rangle$ in $|\Psi_4\rangle$ of Eq.(\ref{qubit-PCG-state}) with $|0\rangle$ and $|d/2-1\rangle$ respectively, one can convert the state $|\Psi_4\rangle$ into $|\Psi_4\rangle^{\prime}$. Moreover, one can check that the algebraic relations for $\{\sigma_x,\sigma_z\}$ and $\{X^{d/2},Z^{d/2}\}$ are the same. Therefore,
if the qubit Pauli operators $\sigma_x$ and $\sigma_z$ are replaced by $X^{d/2}$ and $Z^{d/2}$, the DAVN proof  of Bell  nonlocality induced from $|\Psi_4\rangle$ can be naturally generalized to a DAVN proof associated with $|\Psi_4\rangle^{\prime}$. Since such a generalized proof can be reduced to the qubit version, it is just a trivial result.

By contrast, constructing non-trivial DAVN proofs of Bell nonlocality from qudit non-stabilizer states is far more difficult.
According to the experience of the qubit scenario, some symmetries may considerably simplify the construction. So far, whether there are such proofs for $d=3$ is still unknown. For simplicity, we shall mainly consider the scenario with $d=4$. Accordingly, the qudit Pauli operators $Z=\sum_{k=0}^{3}e^{i\frac{k\pi}{2}}|k\rangle\langle k|$ and $X=\sum_{k=0}^3|k\oplus1\rangle\langle k|$, and
their eigenspectrums are both $1, -1, i$ and $-i$. Similar to the qubit scenario, $Z_1Z_2\cdots Z_n$ can be chosen as an invariant, and the following observation might be helpful in the construction of the desired DAVN proofs.

{\it Observation.---} If $Z_1Z_2\cdots Z_n$ can stabilize a state $|k_1k_2\cdots k_n\rangle$ ($k_{1,2,\cdots,n}=0,1,2,3$), i.e., $Z_1Z_2\cdots Z_n|k_1k_2\cdots k_n\rangle=e^{i\frac{\pi}{2}(k_1+k_2+\cdots+k_n)}|k_1k_2\cdots k_n\rangle=|k_1k_2\cdots k_n\rangle$, one must have $k_1 \oplus k_2 \oplus \cdots\oplus k_n=0$, and vice versa.

Then consider the following quantum state
\begin{widetext}
\begin{align}\label{phi-4}
    |\Psi\rangle_{1234}=&\frac{1}{2\sqrt{14}}(|0000\rangle+|2222\rangle\cr
    &-|0022\rangle+i|0233\rangle-i|2033\rangle+i|0211\rangle-i|2011\rangle+i|1300\rangle-i|3100\rangle-i|1322\rangle+i|3122\rangle\cr
    &-|2002\rangle+i|3023\rangle-i|3203\rangle+i|1021\rangle-i|1201\rangle+i|0130\rangle-i|0310\rangle-i|2132\rangle+i|2312\rangle\cr
    &-|2200\rangle+i|3302\rangle-i|3320\rangle+i|1102\rangle-i|1120\rangle+i|0013\rangle-i|0031\rangle-i|2213\rangle+i|2231\rangle\cr
    &-|0220\rangle+i|2330\rangle-i|0332\rangle+i|2110\rangle-i|0112\rangle+i|3001\rangle-i|1003\rangle-i|3221\rangle+i|1223\rangle\cr
    &-|0202\rangle+i|0323\rangle-i|2303\rangle+i|0121\rangle-i|2101\rangle+i|1030\rangle-i|3010\rangle-i|1232\rangle+i|3212\rangle\cr
    &-|2020\rangle+i|3032\rangle-i|3230\rangle+i|1012\rangle-i|1210\rangle+i|0103\rangle-i|0301\rangle-i|2123\rangle+i|2321\rangle).\cr
    \end{align}
\end{widetext}
Since each component $|k_1k_2k_3k_4\rangle$ in $|\Psi\rangle_{1234}$ satisfies $k_1 \oplus k_2 \oplus k_3 \oplus k_4=0$,  one can get
\begin{align}\label{stabilizer}
Z_1Z_2Z_3Z_4|\Psi\rangle_{1234}=|\Psi\rangle_{1234}
\end{align}
according to the above observation, which indicates that $Z_1Z_2Z_3Z_4$ is a stabilizer of $|\Psi\rangle_{1234}$.
However, $|\Psi\rangle_{1234}$ is NOT a stabilizer state. In fact, for a fully entangled qudit stabilizer state, one can calculate the reduced density matrix on each particle (denote by $\rho_j$ the reduced density matrix on the $j$-th qudit). If $\rho_j=\frac{I}{d}$ holds for any $j$, it is a stabilizer state, and otherwise, it is a non-stabilizer one. It turns out that
$\rho_1=\frac{2}{7}(|0\rangle\langle0|+|2\rangle\langle2|)+\frac{3}{14}(|1\rangle\langle1|+|3\rangle\langle3|)\neq\frac{I}{4}$, and thus $|\Psi\rangle_{1234}$ is a non-stabilizer state.

The stabilizer $Z_1Z_2Z_3Z_4$ can give rise to a joint probability of measuring $Z_1,Z_2,Z_3,Z_4$ on the system of $|\Psi\rangle_{1234}$ \cite{Hermitian-measurement}, which can be described as
\begin{align}\label{joint-probability}
    &P(Z_1=m_1,Z_2=m_2,Z_3=m_3,Z_4=m_4)\cr
=&\left\{
                                         \begin{array}{ll}
                                           \frac{1}{56}, & \hbox{\text{if}~$m_1m_2m_3m_4=1$;} \\
                                           0, & \hbox{otherwise.}
                                         \end{array}
                                       \right.
\end{align}
Here $m_{1,2,3,4}\in\{1,i,-1,-i\}$. This relation can induce a total of $56$ probabilistic all-versus-nothing proofs of Bell nonlocality, and they can be regarded as $56$ qudit Hardy-like quantum pigeonhole (HLQP) paradoxes\cite{Tang2022} (which are essentially special multi-qudit Hardy's paradoxes). Similar to the qubit scenario referred to in Ref.\cite{Tang-DAVN-2022} , combining them together will give rise to a required DAVN proof of Bell nonlocality.
To give a more clear description to these $56$ probabilistic all-versus-nothing proofs, we shall classify them into six types.

\subsection{Type I}

There are two qudit HLQP paradoxes in type I.

Consider  a run of the experiment that the observables $Z_1,Z_2,Z_3$ and $Z_4$ are measured, and the results $Z_1=\alpha,Z_2=\alpha,Z_3=\alpha$ and $Z_4=\alpha$ are obtained, where $\alpha=1$ and $-1$  arise from the contributions of the components $|0000\rangle$ and $|2222\rangle$, respectively. Besides, for
simplicity, hereafter we use the same notation ($Z_i$, $X_j$ etc)
to represent a qudit Pauli operator and the corresponding
output (outcome) in a run of the experiment. Then we can get the following (quantum) properties:
\begin{subequations}\label{Hardy-cond-type-I}
\begin{align}
    Z_1=Z_2=\alpha, & \Longrightarrow~ X_3X_4^3=-i; \label{Hardy-cond-type-I1}\\
    Z_1=Z_3=\alpha, & \Longrightarrow~ X_2X_4^3=-i; \label{Hardy-cond-type-I2}\\
    Z_1=Z_4=\alpha, & \Longrightarrow~ X_2X_3^3=-i; \label{Hardy-cond-type-I3}  \\
    Z_2=Z_3=\alpha, & \Longrightarrow~ X_1X_4^3=i; \label{Hardy-cond-type-I4}  \\
    Z_2=Z_4=\alpha, & \Longrightarrow~ X_1X_3^3=-i; \label{Hardy-cond-type-I5} \\
    Z_3=Z_4=\alpha, & \Longrightarrow~ X_1X_2^3=-i. \label{Hardy-cond-type-I6}
\end{align}
\end{subequations}
Formally, the above relations and the relation $P(Z_1=\alpha,Z_2=\alpha,Z_3=\alpha,Z_4=\alpha)=1/56$ are referred to as Hardy-like conditions. Below we will see that all the involved joint probability relations in the 56 HLQP paradoxes can be described by a unified form of Eq. (\ref{joint-probability}). For convenience, we shall simply refer to the relations of Eqs. (\ref{Hardy-cond-type-I1}-Hardy-cond-type-I6) as Hardy-like conditions (in which although the joint probability relation is not listed, it will automatically be included in derive a HLQP paradox, the same below).

One can also understand the above Hardy-like conditions by post-selections. For example, since we have $Z_1=Z_2=\alpha$ in this run, according to the expression of Eq.(\ref{phi-4}), one can equivalently consider qubit $3$ and $4$ as a post-selected state $(|00\rangle+i|13\rangle-|22\rangle-i|31\rangle)/2$. For other cases ($Z_i=Z_j=\alpha$, $i\neq j$), see Table \ref{TB1}. To give a probabilistic all-versus-nothing proof of Bell nonlocality, apart from using the basic constraints in Table \ref{TB1}, sometimes one can also invoke the extend ones from this table, but note that the basic constraints can lead to the extended ones, while the latter cannot induce the former.

\begin{table*}\caption{The Hardy-like conditions for type I, which are associated with the measurement results $Z_1=\alpha,Z_2=\alpha,Z_3=\alpha$ and $Z_4=\alpha$ (from a post-selection point of view), where the extended constraints are weaker versions of the basic ones.} \label{TB1}
    \centering
    \begin{tabular}{c|ccc}
    \hline
    \hline
        ~~$Z_i,Z_j$ $(i\neq j)$~~ & ~~$|\psi\rangle_{kl}$ ($k\neq l\neq i\neq j$, up to a phase $\pm1$)~~ & ~~ Basic constraints  ~~ &  ~~ Extended constraints ~~\\
  \hline
   ~~$Z_1=Z_2=\alpha$~~  & ~~$|\psi\rangle_{34}=\frac{1}{2}(|00\rangle+i|13\rangle-|22\rangle-i|31\rangle)$~~ &  ~~$X_3X_4^3=-i$~~ & ~~$X_3^2X_4^2=-1$~~ \\
 ~~$Z_1=Z_3=\alpha$~~  & ~~$|\psi\rangle_{24}=\frac{1}{2}(|00\rangle+i|13\rangle-|22\rangle-i|31\rangle)$~~ &  ~~$X_2X_4^3=-i$~~ & ~~$X_2^2X_4^2=-1$~~ \\
 ~~$Z_1=Z_4=\alpha$~~  & ~~$|\psi\rangle_{23}=\frac{1}{2}(|00\rangle+i|13\rangle-|22\rangle-i|31\rangle)$~~ &  ~~$X_2X_3^3=-i$~~ & ~~$X_2^2X_3^2=-1$~~ \\
 ~~$Z_2=Z_3=\alpha$~~  & ~~$|\psi\rangle_{14}=\frac{1}{2}(|00\rangle-i|13\rangle-|22\rangle+i|31\rangle)$~~ &  ~~$X_1X_4^3=i$~~ & ~~$X_4^2X_1^2=-1$~~ \\
 ~~$Z_2=Z_4=\alpha$~~  & ~~$|\psi\rangle_{13}=\frac{1}{2}(|00\rangle+i|13\rangle-|22\rangle-i|31\rangle)$~~ &  ~~$X_1X_3^3=-i$~~ & ~~$X_1^2X_3^2=-1$~~ \\
 ~~$Z_3=Z_4=\alpha$~~  & ~~$|\psi\rangle_{12}=\frac{1}{2}(|00\rangle+i|13\rangle-|22\rangle-i|31\rangle)$~~ &  ~~$X_1X_2^3=-i$~~ & ~~$X_1^2X_2^2=-1$~~ \\
     \hline
   \hline
    \end{tabular}
\end{table*}

Let us show how to construct a HLQP paradox by invoking Eq.(\ref{joint-probability}) and Eqs.(\ref{Hardy-cond-type-I1}-\ref{Hardy-cond-type-I6}). First of all, suppose that four qudits of $|\Psi\rangle_{1234}$ are hold by four observers in different places (hereafter we only consider the space-like separated measurements). Assume that $|\Psi\rangle_{1234}$ admits a local realistic description. As we mentioned, in this run of experiment $Z_1,Z_2,Z_3,Z_4$ are measured and the outcomes $Z_1=Z_2=Z_3=Z_4=\alpha$ are obtained (For each $\alpha$, this happens with a probability of $1/56$). Since we have $Z_1=Z_2=\alpha$, one can infer from Eq.(\ref{Hardy-cond-type-I1}) that if $X_3$ and $X_4^3$ were measured in this run (instead of $Z_3$ and $Z_4$), their outcomes should satisfy $X_3X_4^3=-i$. In fact, according to the assumption of locality, even if other local observables (e.g. $X_1,X_2$) had been measured on qudits $1$ and $2$, one would still get $X_3X_4^3=-i$ (independent of the choice of the measurements on qubit $1$ and $2$). Thus in this run, the outcomes of measuring $X_3$ and $X_4^3$ (determined by the
hidden variables $\lambda$ in the LHV model, and notice that $X_4^3(\lambda)=[X_4(\lambda)]^3$ according to the classical value assignment by LHV model, where $X_4(\lambda)$ and $X_4^3(\lambda)$ stand for the results for measuring $X_4$ and $X_4^3$ respectively) must satisfy $X_3(\lambda)[X_4(\lambda)]^3=-i$. Similar arguments applied to Eqs.(\ref{Hardy-cond-type-I2}-\ref{Hardy-cond-type-I6}) will give rise to
$X_2(\lambda)[X_4(\lambda)]^3=-i,X_2(\lambda)[X_3(\lambda)]^3=-i,X_1(\lambda)[X_4(\lambda)]^3=i,X_1(\lambda)[X_3(\lambda)]^3=-i,
X_1(\lambda)[X_2(\lambda)]^3=-i$. To see the contradiction, one can check the squares of these relations, e.g., the square of $X_3(\lambda)[X_4(\lambda)]^3=-i$ leads to $[X_3(\lambda)]^2[X_4(\lambda)]^2=-1$. Then one have $[X_i(\lambda)]^2[X_j(\lambda)]^2=-1$ ($i\neq j\in\{1,2,3,4\}$). Clearly, these relations cannot hold simultaneously, and thus one can get a qudit HLQP paradox. Besides, notice that $\alpha$ can take two values, indicating that there are two such paradoxes in type I.

Note that in the above argument, one can also get the contradiction based on Eqs.(\ref{Hardy-cond-type-I1}-\ref{Hardy-cond-type-I3}). Since $X_j^2=\pm1~(j\in\{1,2,3,4\})$, the product of the left hand side of  these equations is either $+1$ or $-1$. However, the product of the right hand side is $i$,  a contradiction!

\subsection{Type II}

We will show that  there are six HLQP paradoxes in type II.

To derive any of them, one can consider a run of experiment that the observables $Z_p,Z_q,Z_r$ and $Z_s$ are measured, and the results $Z_p=1,Z_q=1,Z_r=-1$, and $Z_s=-1$ are obtained (which happens with a probability of $1/56$), where
$p\neq q\neq r\neq s\in\{1,2,3,4\}$. Clearly, these results arise from the the contributions of the permutations of the component $|0022\rangle$ (up to a phase factor,the same below), i.e., $|0022\rangle$, $|2002\rangle$, $|2200\rangle$, $|0220\rangle$, $|0202\rangle$ and $|2020\rangle$.
Accordingly, six groups of Hardy-like conditions can be given, and each of  group of them can be associated to one of the above permutated components. Similar to the discussion of type I, these Hardy-like conditions can be derived by a post-selection technique, see Table \ref{TB2}.  Precisely, they can be described as follows.

(i) For $(p,q,r,s)\in\{(1,2,3,4),(3,4,1,2),(1,3,2,4),(2,4,1,3)\}$ (associated withe the components $|0022\rangle$, $|2200\rangle$, $|0202\rangle$ and $|2020\rangle$), their
 Hardy-like conditions are specified by
\begin{subequations}\label{Hardy-cond-type-II}
\begin{align}
    Z_p=1,Z_q=1, & \Longrightarrow~ X_rX_s^3=-i; \label{Hardy-cond-type-II1}\\
    Z_p=1,Z_r=-1, & \Longrightarrow~ X_q^2X_s^2=1; \label{Hardy-cond-type-II2}\\
    Z_p=1,Z_s=-1, & \Longrightarrow~ X_q^2X_r^2=1; \label{Hardy-cond-type-II3}  \\
    Z_q=1,Z_r=-1, & \Longrightarrow~ X_p^2X_s^2=1;\label{Hardy-cond-type-II4}  \\
    Z_q=1,Z_s=-1, & \Longrightarrow~ X_p^2X_r^2=1; \label{Hardy-cond-type-II5} \\
    Z_r=-1,Z_s=-1, & \Longrightarrow~ X_pX_q^3=-i. \label{Hardy-cond-type-II6}
\end{align}
\end{subequations}
\begin{subequations}\label{Hardy-cond-type-II-B}

(ii) For the other two cases, i.e., $(p,q,r,s)\in\{(2,3,1,4),(1,4,2,3)\}$ (associated with the components $|2002\rangle$ and $|0220\rangle$),  let $\alpha=\pm1$, then the corresponding Hardy-like conditions can be written as
\begin{align}
    Z_2=\alpha,Z_3=\alpha, & \Longrightarrow~ X_1X_4^3=i; \label{Hardy-cond-type-II-B1}\\
    Z_2=\alpha,Z_1=-\alpha, & \Longrightarrow~ X_3^2X_4^2=1; \label{Hardy-cond-type-II-B2}\\
    Z_2=\alpha,Z_4=-\alpha, & \Longrightarrow~ X_1^2X_3^2=1; \label{Hardy-cond-type-II-B3}  \\
    Z_3=\alpha,Z_1=-\alpha, & \Longrightarrow~ X_2^2X_4^2=1;\label{Hardy-cond-type-II-B4}  \\
    Z_3=\alpha,Z_4=-\alpha, & \Longrightarrow~ X_1^2X_2^2=1; \label{Hardy-cond-type-II-B5} \\
    Z_1=-\alpha,Z_4=-\alpha, & \Longrightarrow~ X_2X_3^3=-i. \label{Hardy-cond-type-II-B6}
\end{align}
\end{subequations}

\begin{table*}\caption{Six Hardy-like conditions for type II, which are associated with the measurement results $Z_p=1,Z_q=1,Z_r=-1$, and $Z_s=-1$ (from a post-selection point of view).} \label{TB2}
    \centering
    \begin{tabular}{c|ccc}
    \hline
    \hline
        ~~$Z_i,Z_j$ $(i\neq j)$~~ & ~~$|\psi\rangle_{kl}$ ($k\neq l\neq i\neq j$)~~ & ~~ Basic constraints  ~~ &  ~~ Extended constraints ~~\\
  \hline
   ~~$Z_1=1,Z_2=1$~~  & ~~$|\psi\rangle_{34}=\frac{1}{2}(|00\rangle+i|13\rangle-|22\rangle-i|31\rangle)$~~ &  ~~$X_3X_4^3=-i$~~ & ~~$X_3^2X_4^2=-1$~~ \\
 ~~$Z_1=1,Z_3=-1$~~  & ~~$|\psi\rangle_{24}=\frac{1}{2}(-|02\rangle+i|11\rangle-|20\rangle+i|33\rangle)$~~ &  ~~$X_2^2X_4^2=1$~~ & ~~$X_2^2X_4^2=1$~~ \\
 ~~$Z_1=1,Z_4=-1$~~  & ~~$|\psi\rangle_{23}=\frac{1}{2}(-|02\rangle-i|11\rangle-|20\rangle-i|33\rangle)$~~ &  ~~$X_2^2X_3^2=1$~~ & ~~$X_2^2X_3^2=1$~~ \\
 ~~$Z_2=1,Z_3=-1$~~  & ~~$|\psi\rangle_{14}=\frac{1}{2}(-|02\rangle+i|11\rangle-|20\rangle+i|33\rangle)$~~ &  ~~$X_1^2X_4^2=1$~~ & ~~$X_1^2X_4^2=1$~~ \\
 ~~$Z_2=1,Z_4=-1$~~  & ~~$|\psi\rangle_{13}=\frac{1}{2}(-|02\rangle+i|11\rangle-|20\rangle+i|33\rangle)$~~ &  ~~$X_1^2X_3^2=1$~~ & ~~$X_1^2X_3^2=1$~~ \\
 ~~$Z_3=-1,Z_4=-1$~~  & ~~$|\psi\rangle_{12}=\frac{1}{2}(-|00\rangle-i|13\rangle+|22\rangle+i|31\rangle)$~~ &  ~~$X_1X_2^3=-i$~~ & ~~$X_1^2X_2^2=-1$~~ \\
     \hline
   ~~$Z_1=-1,Z_2=1$~~  & ~~$|\psi\rangle_{34}=\frac{1}{2}(-|02\rangle-i|11\rangle-|20\rangle-i|33\rangle)$~~ &  ~~$X_3^2X_4^2=1$~~ & ~~$X_3^2X_4^2=1$~~ \\
 ~~$Z_1=-1,Z_3=1$~~  & ~~$|\psi\rangle_{24}=\frac{1}{2}(-|02\rangle-i|11\rangle-|20\rangle-i|33\rangle)$~~ &  ~~$X_2^2X_4^2=1$~~ & ~~$X_2^2X_4^2=1$~~ \\
 ~~$Z_1=-1,Z_4=-1$~~  & ~~$|\psi\rangle_{23}=\frac{1}{2}(-|00\rangle-i|13\rangle+|22\rangle+i|31\rangle)$~~ &  ~~$X_2X_3^3=-i$~~ & ~~$X_2^2X_3^2=-1$~~ \\
 ~~$Z_2=1,Z_3=1$~~  & ~~$|\psi\rangle_{14}=\frac{1}{2}(|00\rangle-i|13\rangle-|22\rangle+i|31\rangle)$~~ &  ~~$X_1X_4^3=i$~~ & ~~$X_1^2X_4^2=-1$~~ \\
 ~~$Z_2=1,Z_4=-1$~~  & ~~$|\psi\rangle_{13}=\frac{1}{2}(-|02\rangle+i|11\rangle-|20\rangle+i|33\rangle)$~~ &  ~~$X_1^2X_3^2=1$~~ & ~~$X_1^2X_3^2=1$~~ \\
 ~~$Z_3=1,Z_4=-1$~~  & ~~$|\psi\rangle_{12}=\frac{1}{2}(-|02\rangle+i|11\rangle-|20\rangle+i|33\rangle)$~~ &  ~~$X_1^2X_2^2=1$~~ & ~~$X_1^2X_2^2=1$~~ \\
     \hline
   ~~$Z_1=-1,Z_2=-1$~~  & ~~$|\psi\rangle_{34}=\frac{1}{2}(-|00\rangle-i|13\rangle+|22\rangle+i|31\rangle)$~~ &  ~~$X_3X_4^3=-i$~~ & ~~$X_3^2X_4^2=-1$~~ \\
 ~~$Z_1=-1,Z_3=1$~~  & ~~$|\psi\rangle_{24}=\frac{1}{2}(-|02\rangle-i|11\rangle-|20\rangle-i|33\rangle)$~~ &  ~~$X_2^2X_4^2=1$~~ & ~~$X_2^2X_4^2=1$~~ \\
 ~~$Z_1=-1,Z_4=1$~~  & ~~$|\psi\rangle_{23}=\frac{1}{2}(-|02\rangle+i|11\rangle-|20\rangle+i|33\rangle)$~~ &  ~~$X_2^2X_3^2=1$~~ & ~~$X_2^2X_3^2=1$~~ \\
 ~~$Z_2=-1,Z_3=1$~~  & ~~$|\psi\rangle_{14}=\frac{1}{2}(-|02\rangle-i|11\rangle-|20\rangle-i|33\rangle)$~~ &  ~~$X_1^2X_4^2=1$~~ & ~~$X_1^2X_4^2=1$~~ \\
 ~~$Z_2=-1,Z_4=1$~~  & ~~$|\psi\rangle_{13}=\frac{1}{2}(-|02\rangle-i|11\rangle-|20\rangle-i|33\rangle)$~~ &  ~~$X_1^2X_3^2=1$~~ & ~~$X_1^2X_3^2=1$~~ \\
 ~~$Z_3=1,Z_4=1$~~  & ~~$|\psi\rangle_{12}=\frac{1}{2}(|00\rangle+i|13\rangle-|22\rangle-i|31\rangle)$~~ &  ~~$X_1X_2^3=-i$~~ & ~~$X_1^2X_2^2=-1$~~ \\
   \hline
   ~~$Z_1=1,Z_2=-1$~~  & ~~$|\psi\rangle_{34}=\frac{1}{2}(-|02\rangle+i|11\rangle-|20\rangle+i|33\rangle)$~~ &  ~~$X_3^2X_4^2=1$~~ & ~~$X_3^2X_4^2=1$~~ \\
 ~~$Z_1=1,Z_3=-1$~~  & ~~$|\psi\rangle_{24}=\frac{1}{2}(-|02\rangle+i|11\rangle-|20\rangle+i|33\rangle)$~~ &  ~~$X_2^2X_4^2=1$~~ & ~~$X_2^2X_4^2=1$~~ \\
 ~~$Z_1=1,Z_4=1$~~  & ~~$|\psi\rangle_{23}=\frac{1}{2}(|00\rangle+i|13\rangle-|22\rangle-i|31\rangle)$~~ &  ~~$X_2X_3^3=-i$~~ & ~~$X_2^2X_3^2=-1$~~ \\
 ~~$Z_2=-1,Z_3=-1$~~  & ~~$|\psi\rangle_{14}=\frac{1}{2}(-|00\rangle+i|13\rangle+|22\rangle-i|31\rangle)$~~ &  ~~$X_1X_4^3=i$~~ & ~~$X_1^2X_4^2=-1$~~ \\
 ~~$Z_2=-1,Z_4=1$~~  & ~~$|\psi\rangle_{13}=\frac{1}{2}(-|02\rangle-i|11\rangle-|20\rangle-i|33\rangle)$~~ &  ~~$X_1^2X_3^2=1$~~ & ~~$X_1^2X_3^2=1$~~ \\
 ~~$Z_3=-1,Z_4=1$~~  & ~~$|\psi\rangle_{12}=\frac{1}{2}(-|02\rangle-i|11\rangle-|20\rangle-i|33\rangle)$~~ &  ~~$X_1^2X_2^2=1$~~ & ~~$X_1^2X_2^2=1$~~ \\
     \hline
   ~~$Z_1=1,Z_2=-1$~~  & ~~$|\psi\rangle_{34}=\frac{1}{2}(-|02\rangle+i|11\rangle-|20\rangle+i|33\rangle)$~~ &  ~~$X_3^2X_4^2=1$~~ & ~~$X_3^2X_4^2=1$~~ \\
 ~~$Z_1=1,Z_3=1$~~  & ~~$|\psi\rangle_{24}=\frac{1}{2}(|00\rangle+i|13\rangle-|22\rangle-i|31\rangle)$~~ &  ~~$X_2X_4^3=-i$~~ & ~~$X_2^2X_4^2=-1$~~ \\
 ~~$Z_1=1,Z_4=-1$~~  & ~~$|\psi\rangle_{23}=\frac{1}{2}(-|02\rangle-i|11\rangle-|20\rangle-i|33\rangle)$~~ &  ~~$X_2^2X_3^2=1$~~ & ~~$X_2^2X_3^2=1$~~ \\
 ~~$Z_2=-1,Z_3=1$~~  & ~~$|\psi\rangle_{14}=\frac{1}{2}(-|02\rangle-i|11\rangle+|20\rangle-i|33\rangle)$~~ &  ~~$X_1^2X_4^2=1$~~ & ~~$X_1^2X_4^2=1$~~ \\
 ~~$Z_2=-1,Z_4=-1$~~  & ~~$|\psi\rangle_{13}=\frac{1}{2}(-|00\rangle-i|13\rangle+|22\rangle+i|31\rangle)$~~ &  ~~$X_1X_3^3=-i$~~ & ~~$X_1^2X_3^2=-1$~~ \\
 ~~$Z_3=1,Z_4=-1$~~  & ~~$|\psi\rangle_{12}=\frac{1}{2}(-|02\rangle+i|11\rangle-|20\rangle+i|33\rangle)$~~ &  ~~$X_1^2X_2^2=1$~~ & ~~$X_1^2X_2^2=1$~~ \\
     \hline
   ~~$Z_1=-1,Z_2=1$~~  & ~~$|\psi\rangle_{34}=\frac{1}{2}(-|02\rangle-i|11\rangle-|20\rangle-i|33\rangle)$~~ &  ~~$X_3^2X_4^2=1$~~ & ~~$X_3^2X_4^2=1$~~ \\
 ~~$Z_1=-1,Z_3=-1$~~  & ~~$|\psi\rangle_{24}=\frac{1}{2}(-|00\rangle-i|13\rangle+|22\rangle+i|31\rangle)$~~ &  ~~$X_2X_4^3=-i$~~ & ~~$X_2^2X_4^2=-1$~~ \\
 ~~$Z_1=-1,Z_4=1$~~  & ~~$|\psi\rangle_{23}=\frac{1}{2}(-|02\rangle+i|11\rangle-|20\rangle+i|33\rangle)$~~ &  ~~$X_2^2X_3^2=1$~~ & ~~$X_2^2X_3^2=1$~~ \\
 ~~$Z_2=1,Z_3=-1$~~  & ~~$|\psi\rangle_{14}=\frac{1}{2}(-|02\rangle+i|11\rangle+|20\rangle+i|33\rangle)$~~ &  ~~$X_1^2X_4^2=1$~~ & ~~$X_1^2X_4^2=1$~~ \\
 ~~$Z_2=1,Z_4=1$~~  & ~~$|\psi\rangle_{13}=\frac{1}{2}(|00\rangle+i|13\rangle-|22\rangle-i|31\rangle)$~~ &  ~~$X_1X_3^3=-i$~~ & ~~$X_1^2X_3^2=-1$~~ \\
 ~~$Z_3=-1,Z_4=1$~~  & ~~$|\psi\rangle_{12}=\frac{1}{2}(-|02\rangle-i|11\rangle-|20\rangle-i|33\rangle)$~~ &  ~~$X_1^2X_2^2=1$~~ & ~~$X_1^2X_2^2=1$~~ \\
     \hline
 \hline
    \end{tabular}
\end{table*}

First, we show that how to construct a paradoxical argument (HLQP paradox) based on the four groups of Hardy-like conditions in (i).
As  we mentioned above, in this run of experiment, measuring $Z_p,Z_q,Z_r$ and $Z_s$ gives the outcomes $Z_p=1,Z_q=1,Z_r=-1$, and $Z_s=-1$. Assume that $|\Psi\rangle_{1234}$ can be described by a LHV model. Since we have $Z_p=1,Z_q=1$,
one can infer that $X_r(\lambda)[X_s(\lambda)]^3=-i$ by invoking Eq.(\ref{Hardy-cond-type-II1}), which also indicates that $[X_r(\lambda)]^2[X_s(\lambda)]^2=-1$ (note that $X_j^4=1,~j\in\{p,q,r,s\}$). Likewise, other relations  $[X_q(\lambda)]^2[X_s(\lambda)]^2=1$,  $[X_q(\lambda)]^2[X_r(\lambda)]^2=1$,  $[X_p(\lambda)]^2[X_s(\lambda)]^2=1$,  $[X_p(\lambda)]^2[X_r(\lambda)]^2=1$, and $X_p(\lambda)[X_q(\lambda)]^3=-i$ can also be derived from Eqs.(\ref{Hardy-cond-type-II2}-\ref{Hardy-cond-type-II6}) for this run, which leads to a contradiction:  the product of the left hand side of $[X_r(\lambda)]^2[X_s(\lambda)]^2=-1$, $[X_q(\lambda)]^2[X_s(\lambda)]^2=1$, and $[X_q(\lambda)]^2[X_r(\lambda)]^2=1$ is $1$, while the product of the right hand side of them is $-1$. Namely, we have got a qudit HLQP paradox. Since there are four groups of relations in (i), four such HLQP paradoxes can be obtained.

Next, by using similar arguments, one can get the other two HLQP paradoxes based on the two groups of Hardy-like conditions in (ii).
Therefore,  type II contains a total of six HLQP paradoxes (there is a correspondence between these paradoxes and the components $|0022\rangle$, $|2002\rangle$, $|2200\rangle$, $|0220\rangle$, $|0202\rangle$ and $|2020\rangle$).

\subsection{Type III}

As mentioned above, each HLQP paradox  in type II can be associated to a permutation of the component $|0022\rangle$. Likewise,  HLQP paradoxes in type III are associated with $12$ permutations  of the component $|0033\rangle$ (up to a phase $i$ or $-i$). Therefore,
type III contains a total of $12$ HLQP paradoxes.

To show that, similar to the discussions from types I and II, assume that in a
run of the experiment, measuring $Z_p,Z_q,Z_r,Z_s$ give rise to the
results  $Z_p=\alpha,Z_q=-\alpha,Z_r=-i$, $Z_s=-i$, where $p\neq q\neq r\neq s\in\{1,2,3,4\}$ and $\alpha=\pm1$. Clearly, there are $12$ groups of such results, which arise from the contributions of $12$ permutations of the component $|0233\rangle$ (up to a phase).  Accordingly, 12 groups of Hardy-like conditions can be derived.
For simplicity, we group them into two sub-classes:
type III-A and type III-B, see Tables \ref{TB3} and \ref{TB4}. One can  invoke part of (or all of) the relations in each group of the Hardy-like conditions, and use similar techniques referred to in types I and II, to construct a HLQP paradox.

Note that in Tables \ref{TB3} and \ref{TB4}, the 12 groups of Hardy-like conditions can be sorted into two families. Apart from the default joint probability relation (see the aforementioned statement),  each group in the first family contains six constraints (e.g. the one associated with the term $i|0233\rangle$), while in the second family, each group of Hardy-like conditions only contains four constraints (e.g. the one associated with the term $i|3023\rangle$). Next, we shall give two examples to show how to induce the HLQP paradoxes for both families.

In the first example, let us consider the case $p=1,q=2,r=3,s=4$, namely, in the run of the experiment, the results of measuring $Z_1,Z_2,Z_3,Z_4$ are $1,-1,-i,-i$ respectively (which happens with a probability of $1/56$).  Accordingly,
the Hardy-like conditions are associated with the term $i|0233\rangle$, which can be represented as follows (see Tables \ref{TB3}):
\begin{subequations}\label{Hardy-cond-type-III-1}
\begin{align}
    Z_1=1,Z_2=-1, & \Longrightarrow~ X_3^2X_4^2=1; \label{Hardy-cond-type-III1}\\
    Z_1=1,Z_3=-i, & \Longrightarrow~ X_2^2X_4^2=-1; \label{Hardy-cond-type-III2}\\
    Z_1=1,Z_4=-i, & \Longrightarrow~ X_2X_3^3=1; \label{Hardy-cond-type-III3}  \\
    Z_2=-1,Z_3=-i, & \Longrightarrow~ X_1X_4^3=-1;\label{Hardy-cond-type-III4}  \\
    Z_2=-1,Z_4=-i, & \Longrightarrow~ X_1^2X_3^2=-1; \label{Hardy-cond-type-III5} \\
    Z_3=-i,Z_4=-i, & \Longrightarrow~ X_1^2X_2^2=-1. \label{Hardy-cond-type-III6}
\end{align}
\end{subequations}
Analogous to the arguments of type I and II, assume that the state $|\Psi\rangle_{1234}$ admits a local realistic description. Since one have $Z_1=1,Z_2=-1$, according to  Eq.(\ref{Hardy-cond-type-III1}),  if $X_3^2$ and $X_4^2$ were measured in this run, the results should satisfy $X_3^2X_4^2=1$. Note that this constraint is even independent of the choice for the measurements on qubit $1$ and $2$ (the assumption of locality by the LHV model). Namely, if one measure $X_3^2$ and $X_4^2$ in this run, one can infer that the corresponding outcomes must satisfy
$X_3^2(\lambda)X_4^2(\lambda)=[X_3(\lambda)]^2[X_4{\lambda}]^2=1$. Likewise, one can conclude from  Eqs.(\ref{Hardy-cond-type-III2}-\ref{Hardy-cond-type-III3})
that $[X_2(\lambda)]^2[X_4{\lambda}]^2=-1$, and $X_2(\lambda)[X_3{\lambda}]^3=1$, where $X_2(\lambda)[X_3{\lambda}]^3=1$ also implies that
$[X_2(\lambda)]^2[X_3{\lambda}]^2=1$. However, the products of both sides of the other two relations, $[X_3(\lambda)]^2[X_4{\lambda}]^2=1$ and $[X_2(\lambda)]^2[X_4{\lambda}]^2=-1$, gives rise to $[X_2(\lambda)]^2[X_3{\lambda}]^2=-1$, a contradiction. Then one have get a HLQP paradox.

Note that here we only  need to invoke part of the Hardy-like conditions, i.e.,  Eqs.(\ref{Hardy-cond-type-III1}-\ref{Hardy-cond-type-III3}). Similar HLQP paradoxes can be induced from the other three groups of Hardy-like conditions in the first family.

The second example corresponds to the case that the outcomes $Z_1=-i,Z_2=1,Z_3=-1,Z_4=-i$ are obtained in a run of the experiment (which also happens with a probability of $1/56$). Then this group of Hardy-like conditions (associated with the term $i|3023\rangle$, which belong to the second family, also see Tables \ref{TB3}) can be described as
\begin{subequations}\label{Hardy-cond-type-III-A}
\begin{align}
    Z_1=-i,Z_3=-1, & \Longrightarrow~ X_2^2X_4^2=-1; \label{Hardy-cond-type-III-A1}\\
    Z_1=-i,Z_4=-i, & \Longrightarrow~ X_2^2X_3^2=-1; \label{Hardy-cond-type-III-A2}  \\
    Z_2=1,Z_3=-1, & \Longrightarrow~ X_1^2X_4^2=1;\label{Hardy-cond-type-III-A3}  \\
    Z_2=1,Z_4=-i, & \Longrightarrow~ X_1^2X_3^2=-1. \label{Hardy-cond-type-III-A4}
\end{align}
\end{subequations}
Likewise, based on  the Hardy-like conditions of Eqs.(\ref{Hardy-cond-type-III-A1}-\ref{Hardy-cond-type-III-A4}), one can infer that the relations $[X_2(\lambda)]^2[X_4{\lambda}]^2=-1$, $[X_2(\lambda)]^2[X_3{\lambda}]^2=-1$, $[X_1(\lambda)]^2[X_4{\lambda}]^2=1$, and $[X_1(\lambda)]^2[X_3{\lambda}]^2=-1$ must be hold simultaneously according to the LHV model. However, the products of both sides of these relations lead to a contradiction that $1=-1$. Therefore, this group of Hardy-like conditions can induce a HLQP paradox as well.

For other groups of the Hardy-like conditions from the second family, by using the same method, one can derive the other $7$ HLQP paradoxes.
But note that sometimes one should turn the basic constraints to the extended ones (see Table \ref{TB3} and \ref{TB4}), then get the contradiction from the latter.

In conclusion, one can indeed construct $12$ HLQP paradoxes for the type.

\begin{table*}\caption{Six groups of Hardy-like conditions for type III-A, in which the measurement results for $Z_1$, $Z_2$, $Z_3$, $Z_4$ arise from the contributions of the terms $i|0233\rangle$, $i|3023\rangle$, $i|3302\rangle$, $i|2330\rangle$, $i|0323\rangle$, $i|3032\rangle$, respectively. If the post-selected state is not an eigenstate of $X_i^uX_j^v$ ($u,v\in\{1,2,3\}$), there is no deterministic value constraints for $X_i^u$ and $X_j^v$, and we use the notation ``---" to represent that (the same below).} \label{TB3}
    \centering
    \centering
    \begin{tabular}{c|ccc}
    \hline
    \hline
        ~~$Z_i,Z_j$ $(i\neq j)$~~ & ~~$|\psi\rangle_{kl}$ ($k\neq l\neq i\neq j$)~~ & ~~ Basic constraints  ~~  &  ~~ Extended constraints ~~\\
  \hline
 ~~$Z_1=1,Z_2=-1$~~  & ~~$|\psi\rangle_{34}=\frac{1}{2}(-|02\rangle+i|11\rangle-|20\rangle+i|33\rangle)$~~ &  ~~$X_3^2X_4^2=1$~~ & ~~$X_3^2X_4^2=1$~~\\
 ~~$Z_1=1,Z_3=-i$~~  & ~~$|\psi\rangle_{24}=\frac{1}{2}(-i|01\rangle+i|10\rangle+i|23\rangle-i|32\rangle)$~~ &  ~~$X_2^2X_4^2=-1$~~ & ~~$X_2^2X_4^2=-1$~~ \\
 ~~$Z_1=1,Z_4=-i$~~  & ~~$|\psi\rangle_{23}=\frac{1}{2}(i|01\rangle+i|10\rangle+i|23\rangle+i|32\rangle)$~~ &  ~~$X_2X_3^3=1$~~ & ~~$X_2^2X_3^2=1$~~ \\
 ~~$Z_2=-1,Z_3=-i$~~  & ~~$|\psi\rangle_{14}=\frac{1}{2}(i|03\rangle-i|12\rangle+i|21\rangle-i|30\rangle)$~~ &  ~~$X_1X_4^3=-1$~~ &  ~~$X_1^2X_4^2=1$~~\\
 ~~$Z_2=-1,Z_4=-i$~~  & ~~$|\psi\rangle_{13}=\frac{1}{2}(i|03\rangle+i|12\rangle-i|21\rangle-i|30\rangle)$~~ &  ~~$X_1^2X_3^2=-1$~~ & ~~$X_1^2X_3^2=-1$~~ \\
 ~~$Z_3=-i,Z_4=-i$~~  & ~~$|\psi\rangle_{12}=\frac{1}{\sqrt{2}}(i|02\rangle-i|20\rangle)$~~ &  ~~$X_1^2X_2^2=-1$~~ & ~~$X_1^2X_2^2=-1$~~ \\
     \hline
 ~~$Z_1=-i,Z_2=1$~~  & ~~$|\psi\rangle_{34}=\frac{1}{2}(i|01\rangle-i|10\rangle+i|23\rangle+i|32\rangle)$~~ &  ~~$\text{---}$~~ & ~~$\text{---}$~~\\
 ~~$Z_1=-i,Z_3=-1$~~  & ~~$|\psi\rangle_{24}=\frac{1}{2}(i|03\rangle+i|12\rangle-i|21\rangle-i|30\rangle)$~~ &  ~~$X_2^2X_4^2=-1$~~ & ~~$X_2^2X_4^2=-1$~~ \\
 ~~$Z_1=-i,Z_4=-i$~~  & ~~$|\psi\rangle_{23}=\frac{1}{\sqrt{2}}(i|02\rangle-i|20\rangle)$~~ &  ~~$X_2^2X_3^2=-1$~~ & ~~$X_2^2X_3^2=-1$~~ \\
 ~~$Z_2=1,Z_3=-1$~~  & ~~$|\psi\rangle_{14}=\frac{1}{2}(-|02\rangle+i|11\rangle-|20\rangle+i|33\rangle)$~~ &  ~~$X_1^2X_4^2=1$~~ &  ~~$X_1^2X_4^2=1$~~\\
 ~~$Z_2=1,Z_4=-i$~~  & ~~$|\psi\rangle_{13}=\frac{1}{2}(i|01\rangle-i|10\rangle-i|23\rangle+i|32\rangle)$~~ &  ~~$X_1^2X_3^2=-1$~~ & ~~$X_1^2X_3^2=-1$~~ \\
 ~~$Z_3=-1,Z_4=-i$~~  & ~~$|\psi\rangle_{12}=\frac{1}{2}(i|03\rangle+i|12\rangle-i|21\rangle+i|30\rangle)$~~ &  ~~$\text{---}$~~ & ~~$\text{---}$~~ \\
     \hline
 ~~$Z_1=-i,Z_2=-i$~~  & ~~$|\psi\rangle_{34}=\frac{1}{\sqrt{2}}(i|02\rangle-i|20\rangle)$~~ &  ~~$X_3^2X_4^2=-1$~~ & ~~$X_3^2X_4^2=-1$~~\\
 ~~$Z_1=-i,Z_3=1$~~  & ~~$|\psi\rangle_{24}=\frac{1}{2}(i|01\rangle-i|10\rangle-i|23\rangle+i|32\rangle)$~~ &  ~~$X_2^2X_4^2=-1$~~ & ~~$X_2^2X_4^2=-1$~~ \\
 ~~$Z_1=-i,Z_4=-1$~~  & ~~$|\psi\rangle_{23}=\frac{1}{2}(i|03\rangle+i|12\rangle+i|21\rangle+i|30\rangle)$~~ &  ~~$X_2X_3^3=1$~~ & ~~$X_2^2X_3^2=1$~~ \\
 ~~$Z_2=-i,Z_3=1$~~  & ~~$|\psi\rangle_{14}=\frac{1}{2}(-i|01\rangle+i|10\rangle-i|23\rangle+i|32\rangle)$~~ &  ~~$X_1X_4^3=-1$~~ &  ~~$X_1^2X_4^2=1$~~\\
 ~~$Z_2=-i,Z_4=-1$~~  & ~~$|\psi\rangle_{13}=\frac{1}{2}(-i|03\rangle-i|12\rangle+i|21\rangle+i|30\rangle)$~~ &  ~~$X_1^2X_3^2=-1$~~ & ~~$X_1^2X_3^2=-1$~~ \\
 ~~$Z_3=1,Z_4=-1$~~  & ~~$|\psi\rangle_{12}=\frac{1}{2}(-|02\rangle+i|11\rangle-|20\rangle+i|33\rangle)$~~ &  ~~$X_1^2X_2^2=1$~~ & ~~$X_1^2X_2^2=1$~~ \\
     \hline
 ~~$Z_1=-1,Z_2=-i$~~  & ~~$|\psi\rangle_{34}=\frac{1}{2}(-i|03\rangle+i|12\rangle+i|21\rangle+i|30\rangle)$~~ &  ~~$\text{---}$~~ & ~~$\text{---}$~~\\
 ~~$Z_1=-1,Z_3=-i$~~  & ~~$|\psi\rangle_{24}=\frac{1}{2}(-i|03\rangle-i|12\rangle+i|21\rangle+i|30\rangle)$~~ &  ~~$X_2^2X_4^2=-1$~~ & ~~$X_2^2X_4^2=-1$~~ \\
 ~~$Z_1=-1,Z_4=1$~~  & ~~$|\psi\rangle_{23}=\frac{1}{2}(-|02\rangle+i|11\rangle-|20\rangle+i|33\rangle)$~~ &  ~~$X_2^2X_3^2=1$~~ & ~~$X_2^2X_3^2=1$~~ \\
 ~~$Z_2=-i,Z_3=-i$~~  & ~~$|\psi\rangle_{14}=\frac{1}{\sqrt{2}}(-i|02\rangle+i|20\rangle)$~~ &  ~~$X_1^2X_4^2=-1$~~ &  ~~$X_1^2X_4^2=-1$~~\\
 ~~$Z_2=-i,Z_4=1$~~  & ~~$|\psi\rangle_{13}=\frac{1}{2}(-i|01\rangle+i|10\rangle+i|23\rangle-i|32\rangle)$~~ &  ~~$X_1^2X_3^2=-1$~~ & ~~$X_1^2X_3^2=-1$~~ \\
 ~~$Z_3=-i,Z_4=1$~~  & ~~$|\psi\rangle_{12}=\frac{1}{2}(i|01\rangle+i|10\rangle+i|23\rangle-i|32\rangle)$~~ & ~~$\text{---}$~~ & ~~$\text{---}$~~\\
     \hline
 ~~$Z_1=1,Z_2=-i$~~  & ~~$|\psi\rangle_{34}=\frac{1}{2}(-i|01\rangle-i|10\rangle+i|23\rangle-i|32\rangle)$~~ &  ~~$\text{---}$~~ & ~~$\text{---}$~~\\
 ~~$Z_1=1,Z_3=-1$~~  & ~~$|\psi\rangle_{24}=\frac{1}{2}(-|02\rangle+i|11\rangle-|20\rangle+i|33\rangle)$~~ &  ~~$X_2^2X_4^2=1$~~ & ~~$X_2^2X_4^2=1$~~ \\
 ~~$Z_1=1,Z_4=-i$~~  & ~~$|\psi\rangle_{23}=\frac{1}{2}(i|01\rangle+i|10\rangle+i|23\rangle+i|32\rangle)$~~ &  ~~$X_2X_3^3=1$~~ & ~~$X_2^2X_3^2=1$~~ \\
 ~~$Z_2=-i,Z_3=-1$~~  & ~~$|\psi\rangle_{14}=\frac{1}{2}(i|03\rangle-i|12\rangle+i|21\rangle-i|30\rangle)$~~ &  ~~$X_1X_4^3=-1$~~ &  ~~$X_1^2X_4^2=1$~~\\
 ~~$Z_2=-i,Z_4=-i$~~  & ~~$|\psi\rangle_{13}=\frac{1}{\sqrt{2}}(i|02\rangle-i|20\rangle)$~~ &  ~~$X_1^2X_3^2=-1$~~ & ~~$X_1^2X_3^2=-1$~~ \\
 ~~$Z_3=-1,Z_4=-i$~~  & ~~$|\psi\rangle_{12}=\frac{1}{2}(i|03\rangle+i|12\rangle-i|21\rangle+i|30\rangle)$~~ &  ~~$\text{---}$~~ & ~~$\text{---}$~~ \\
     \hline
 ~~$Z_1=-i,Z_2=1$~~  & ~~$|\psi\rangle_{34}=\frac{1}{2}(i|01\rangle-i|10\rangle+i|23\rangle+i|32\rangle)$~~ & ~~$\text{---}$~~ & ~~$\text{---}$~~\\
 ~~$Z_1=-i,Z_3=-i$~~  & ~~$|\psi\rangle_{24}=\frac{1}{\sqrt{2}}(i|02\rangle-i|20\rangle)$~~ &  ~~$X_2^2X_4^2=-1$~~ & ~~$X_2^2X_4^2=-1$~~ \\
 ~~$Z_1=-i,Z_4=-1$~~  & ~~$|\psi\rangle_{23}=\frac{1}{2}(i|03\rangle+i|12\rangle+i|21\rangle+i|30\rangle)$~~ &  ~~$X_2X_3^3=1$~~ & ~~$X_2^2X_3^2=1$~~ \\
 ~~$Z_2=1,Z_3=-i$~~  & ~~$|\psi\rangle_{14}=\frac{1}{2}(-i|01\rangle+i|10\rangle-i|23\rangle+i|32\rangle)$~~ &  ~~$X_1X_4^3=-1$~~ &  ~~$X_1^2X_4^2=1$~~\\
 ~~$Z_2=1,Z_4=-1$~~  & ~~$|\psi\rangle_{13}=\frac{1}{2}(-|02\rangle+i|11\rangle-|20\rangle+i|33\rangle)$~~ &  ~~$X_1^2X_3^2=1$~~ & ~~$X_1^2X_3^2=1$~~ \\
 ~~$Z_3=-i,Z_4=-1$~~  & ~~$|\psi\rangle_{12}=\frac{1}{2}(-i|03\rangle-i|12\rangle-i|21\rangle+i|30\rangle)$~~ &  ~~$\text{---}$~~ & ~~$\text{---}$~~ \\
     \hline
 \hline
    \end{tabular}
\end{table*}

\begin{table*}\caption{Six groups of Hardy-like conditions for type III-B, in which the measurement results for $Z_1$, $Z_2$, $Z_3$, $Z_4$ arise from the contributions of the terms $-i|2033\rangle$, $-i|3203\rangle$, $-i|3320\rangle$, $-i|0332\rangle$, $-i|2303\rangle$, $-i|3230\rangle$, respectively.} \label{TB4}
    \centering
    \centering
    \begin{tabular}{c|ccc}
    \hline
    \hline
        ~~$Z_i,Z_j$ $(i\neq j)$~~ & ~~$|\psi\rangle_{kl}$ ($k\neq l\neq i\neq j$)~~ & ~~ Basic constraints  ~~  &  ~~ Extended constraints ~~\\
  \hline
 ~~$Z_1=-1,Z_2=1$~~  & ~~$|\psi\rangle_{34}=\frac{1}{2}(-|02\rangle-i|11\rangle-|20\rangle-i|33\rangle)$~~ &  ~~$X_3^2X_4^2=1$~~ & ~~$X_3^2X_4^2=1$~~\\
 ~~$Z_1=-1,Z_3=-i$~~  & ~~$|\psi\rangle_{24}=\frac{1}{2}(-i|03\rangle-i|12\rangle+i|21\rangle+i|30\rangle)$~~ &  ~~$X_2^2X_4^2=-1$~~ & ~~$X_2^2X_4^2=-1$~~ \\
 ~~$Z_1=-1,Z_4=-i$~~  & ~~$|\psi\rangle_{23}=\frac{1}{2}(-i|03\rangle-i|12\rangle-i|21\rangle-i|30\rangle)$~~ &  ~~$X_2X_3^3=1$~~ & ~~$X_2^2X_3^2=1$~~ \\
 ~~$Z_2=1,Z_3=-i$~~  & ~~$|\psi\rangle_{14}=\frac{1}{2}(-i|01\rangle+i|10\rangle-i|23\rangle+i|32\rangle)$~~ &  ~~$X_1X_4^3=-1$~~ &  ~~$X_1^2X_4^2=1$~~\\
 ~~$Z_2=1,Z_4=-i$~~  & ~~$|\psi\rangle_{13}=\frac{1}{2}(i|01\rangle-i|10\rangle-i|23\rangle+i|32\rangle)$~~ &  ~~$X_1^2X_3^2=-1$~~ & ~~$X_1^2X_3^2=-1$~~ \\
 ~~$Z_3=-i,Z_4=-i$~~  & ~~$|\psi\rangle_{12}=\frac{1}{\sqrt{2}}(i|02\rangle-i|20\rangle)$~~ &  ~~$X_1^2X_2^2=-1$~~ & ~~$X_1^2X_2^2=-1$~~ \\
     \hline
 ~~$Z_1=-i,Z_2=-1$~~  & ~~$|\psi\rangle_{34}=\frac{1}{2}(-i|03\rangle+i|12\rangle-i|21\rangle-i|30\rangle)$~~ &  ~~$\text{---}$~~ & ~~$\text{---}$~~\\
 ~~$Z_1=-i,Z_3=1$~~  & ~~$|\psi\rangle_{24}=\frac{1}{2}(i|01\rangle-i|10\rangle-i|23\rangle+i|32\rangle)$~~ &  ~~$X_2^2X_4^2=-1$~~ & ~~$X_2^2X_4^2=-1$~~ \\
 ~~$Z_1=-i,Z_4=-i$~~  & ~~$|\psi\rangle_{23}=\frac{1}{\sqrt{2}}(i|02\rangle-i|20\rangle)$~~ &  ~~$X_2^2X_3^2=-1$~~ & ~~$X_2^2X_3^2=-1$~~ \\
 ~~$Z_2=-1,Z_3=1$~~  & ~~$|\psi\rangle_{14}=\frac{1}{2}(-|02\rangle-i|11\rangle-|20\rangle-i|33\rangle)$~~ &  ~~$X_1^2X_4^2=1$~~ &  ~~$X_1^2X_4^2=1$~~\\
 ~~$Z_2=-1,Z_4=-i$~~  & ~~$|\psi\rangle_{13}=\frac{1}{2}(i|03\rangle+i|12\rangle-i|21\rangle-i|30\rangle)$~~ &  ~~$X_1^2X_3^2=-1$~~ & ~~$X_1^2X_3^2=-1$~~ \\
 ~~$Z_3=1,Z_4=-i$~~  & ~~$|\psi\rangle_{12}=\frac{1}{2}(i|01\rangle-i|10\rangle-i|23\rangle-i|32\rangle)$~~ &  ~~$\text{---}$~~ & ~~$\text{---}$~~ \\
     \hline
 ~~$Z_1=-i,Z_2=-i$~~  & ~~$|\psi\rangle_{34}=\frac{1}{\sqrt{2}}(i|02\rangle-i|20\rangle)$~~ &  ~~$X_3^2X_4^2=-1$~~ & ~~$X_3^2X_4^2=-1$~~\\
 ~~$Z_1=-i,Z_3=-1$~~  & ~~$|\psi\rangle_{24}=\frac{1}{2}(i|03\rangle+i|12\rangle-i|21\rangle-i|30\rangle)$~~ &  ~~$X_2^2X_4^2=-1$~~ & ~~$X_2^2X_4^2=-1$~~ \\
 ~~$Z_1=-i,Z_4=1$~~  & ~~$|\psi\rangle_{23}=\frac{1}{2}(-i|01\rangle-i|10\rangle-i|23\rangle-i|32\rangle)$~~ &  ~~$X_2X_3^3=1$~~ & ~~$X_2^2X_3^2=1$~~ \\
 ~~$Z_2=-i,Z_3=-1$~~  & ~~$|\psi\rangle_{14}=\frac{1}{2}(i|03\rangle-i|12\rangle+i|21\rangle-i|30\rangle)$~~ &  ~~$X_1X_4^3=-1$~~ &  ~~$X_1^2X_4^2=1$~~\\
 ~~$Z_2=-i,Z_4=1$~~  & ~~$|\psi\rangle_{13}=\frac{1}{2}(-i|01\rangle+i|10\rangle+i|23\rangle-i|32\rangle)$~~ &  ~~$X_1^2X_3^2=-1$~~ & ~~$X_1^2X_3^2=-1$~~ \\
 ~~$Z_3=-1,Z_4=1$~~  & ~~$|\psi\rangle_{12}=\frac{1}{2}(-|02\rangle-i|11\rangle-|20\rangle-i|33\rangle)$~~ &  ~~$X_1^2X_2^2=1$~~ & ~~$X_1^2X_2^2=1$~~ \\
     \hline
 ~~$Z_1=1,Z_2=-i$~~  & ~~$|\psi\rangle_{34}=\frac{1}{2}(-i|01\rangle-i|10\rangle+i|23\rangle-i|32\rangle)$~~ &  ~~$\text{---}$~~ & ~~$\text{---}$~~\\
 ~~$Z_1=1,Z_3=-i$~~  & ~~$|\psi\rangle_{24}=\frac{1}{2}(-i|01\rangle+i|10\rangle+i|23\rangle-i|32\rangle)$~~ &  ~~$X_2^2X_4^2=-1$~~ & ~~$X_2^2X_4^2=-1$~~ \\
 ~~$Z_1=1,Z_4=-1$~~  & ~~$|\psi\rangle_{23}=\frac{1}{2}(-|02\rangle-i|11\rangle-|20\rangle-i|33\rangle)$~~ &  ~~$X_2^2X_3^2=1$~~ & ~~$X_2^2X_3^2=1$~~ \\
 ~~$Z_2=-i,Z_3=-i$~~  & ~~$|\psi\rangle_{14}=\frac{1}{\sqrt{2}}(-i|02\rangle+i|20\rangle)$~~ &  ~~$X_1^2X_4^2=-1$~~ &  ~~$X_1^2X_4^2=-1$~~\\
 ~~$Z_2=-i,Z_4=-1$~~  & ~~$|\psi\rangle_{13}=\frac{1}{2}(-i|03\rangle-i|12\rangle+i|21\rangle+i|30\rangle)$~~ &  ~~$X_1^2X_3^2=-1$~~ & ~~$X_1^2X_3^2=-1$~~ \\
 ~~$Z_3=-i,Z_4=-1$~~  & ~~$|\psi\rangle_{12}=\frac{1}{2}(-i|03\rangle-i|12\rangle-i|21\rangle+i|30\rangle)$~~ & ~~$\text{---}$~~ & ~~$\text{---}$~~\\
     \hline
 ~~$Z_1=-1,Z_2=-i$~~  & ~~$|\psi\rangle_{34}=\frac{1}{2}(-i|03\rangle+i|12\rangle+i|21\rangle+i|30\rangle)$~~ &  ~~$\text{---}$~~ & ~~$\text{---}$~~\\
 ~~$Z_1=-1,Z_3=1$~~  & ~~$|\psi\rangle_{24}=\frac{1}{2}(-|02\rangle-i|11\rangle-|20\rangle-i|33\rangle)$~~ &  ~~$X_2^2X_4^2=1$~~ & ~~$X_2^2X_4^2=1$~~ \\
 ~~$Z_1=-1,Z_4=-i$~~  & ~~$|\psi\rangle_{23}=\frac{1}{2}(-i|03\rangle-i|12\rangle-i|21\rangle-i|30\rangle)$~~ &  ~~$X_2X_3^3=1$~~ & ~~$X_2^2X_3^2=1$~~ \\
 ~~$Z_2=-i,Z_3=1$~~  & ~~$|\psi\rangle_{14}=\frac{1}{2}(-i|01\rangle+i|10\rangle-i|23\rangle+i|32\rangle)$~~ &  ~~$X_1X_4^3=-1$~~ &  ~~$X_1^2X_4^2=1$~~\\
 ~~$Z_2=-i,Z_4=-i$~~  & ~~$|\psi\rangle_{13}=\frac{1}{\sqrt{2}}(i|02\rangle-i|20\rangle)$~~ &  ~~$X_1^2X_3^2=-1$~~ & ~~$X_1^2X_3^2=-1$~~ \\
 ~~$Z_3=1,Z_4=-i$~~  & ~~$|\psi\rangle_{12}=\frac{1}{2}(i|01\rangle-i|10\rangle-i|23\rangle-i|32\rangle)$~~ &  ~~$\text{---}$~~ & ~~$\text{---}$~~ \\
     \hline
 ~~$Z_1=-i,Z_2=-1$~~  & ~~$|\psi\rangle_{34}=\frac{1}{2}(-i|03\rangle+i|12\rangle-i|21\rangle-i|30\rangle)$~~ & ~~$\text{---}$~~ & ~~$\text{---}$~~\\
 ~~$Z_1=-i,Z_3=-i$~~  & ~~$|\psi\rangle_{24}=\frac{1}{\sqrt{2}}(i|02\rangle-i|20\rangle)$~~ &  ~~$X_2^2X_4^2=-1$~~ & ~~$X_2^2X_4^2=-1$~~ \\
 ~~$Z_1=-i,Z_4=1$~~  & ~~$|\psi\rangle_{23}=\frac{1}{2}(-i|01\rangle-i|10\rangle-i|23\rangle-i|32\rangle)$~~ &  ~~$X_2X_3^3=1$~~ & ~~$X_2^2X_3^2=1$~~ \\
 ~~$Z_2=-1,Z_3=-i$~~  & ~~$|\psi\rangle_{14}=\frac{1}{2}(i|03\rangle-i|12\rangle+i|21\rangle-i|30\rangle)$~~ &  ~~$X_1X_4^3=-1$~~ &  ~~$X_1^2X_4^2=1$~~\\
 ~~$Z_2=-1,Z_4=1$~~  & ~~$|\psi\rangle_{13}=\frac{1}{2}(-|02\rangle-i|11\rangle-|20\rangle-i|33\rangle)$~~ &  ~~$X_1^2X_3^2=1$~~ & ~~$X_1^2X_3^2=1$~~ \\
 ~~$Z_3=-i,Z_4=1$~~  & ~~$|\psi\rangle_{12}=\frac{1}{2}(i|01\rangle+i|10\rangle+i|23\rangle-i|32\rangle)$~~ &  ~~$\text{---}$~~ & ~~$\text{---}$~~ \\
     \hline
 \hline
    \end{tabular}
\end{table*}

\subsection{Types IV,V, and VI}

The HLQP paradoxes in types IV, V and VI are associated with  permutations  of the components $|0211\rangle$,
$|1300\rangle$  and $|1322\rangle$ respectively (up to a phase $i$ or $-i$).  Note that each of these components has $12$ permutations,
and accordingly, one can find $12$ HLQP paradoxes in each of above types. Besides, since the forms of the permutations in each of these types are very close to those in type III, one may infer that the Hardy-like conditions for types IV, V and VI also have similar structures.
This is indeed the case.  To see that, one can check them in Appendix A, in which the detailed Hardy-like conditions are listed in
Tables \ref{TB5}-\ref{TB10}.

Clearly, similar arguments in type III  also apply to types IV, V, and VI, i.e.,  one can construct 12 HLQP paradoxes for each of these types.

\subsection{The DAVN proof}

According to the above discussion,  a total of $36$ qudit HLQP paradoxes in types III, IV, V, and VI can be derived.
Also notice that types I, II  and III can induce $2$, $6$ and $12$ HLQP paradoxes respectively. Namely, we can get $56$ HLQP paradoxes from the non-stabilizer states $|\Psi\rangle_{1234}$. Combine the $56$ paradoxes together, one can get a  DAVN proof for Bell nonlocality.  This is because, in any run of the experiment, measuring $Z_1,Z_2,Z_3$ and $Z_4$ can always get one of the $56$ groups of the results, i.e., $Z_1=e^{ik_1\frac{\pi}{2}},Z_2=e^{ik_2\frac{\pi}{2}},Z_3=e^{ik_3\frac{\pi}{2}}$ and $Z_4=e^{ik_4\frac{\pi}{2}}$, which arise from the contribution of one of the 56 components, $|k_1k_2k_3k_4\rangle$, in $|\Psi\rangle_{1234}$. As we have discussed above, for each group of the results, we can always construct a paradoxical argument, indicating that the combined proof for Bell nonlocality the is  deterministic (which can rule out the LHV model with a success probability of $100\%$).

In fact, the difficulty of construct such a DAVN proof for Bell nonlocality mainly arises from how to choose a proper non-stabilizer state. Precisely, since we have used the observation to simplify the construction, i.e., the components chosen in the state should satisfy some specific conditions (e.g. each component $|k_1k_2k_3k_4\rangle$ needs to satisfy $k_1\oplus k_2\oplus k_3\oplus k_4=0$), the remaining problem is how to arrange the phase factors (or coefficients) of these components. To get enough Hardy-like conditions so that enough HLQP paradoxes can be produced, one should ensure that, most of (or sometimes even all of) the post-selected states produced by measuring $Z_i$ and $Z_j$ be eigenstates of $X_kX_l$ ($i\neq j\neq k\neq l$). This requirement make the arrangement of the phase factors rather difficult. There is no doubt that there exist other non-stabilizer states which can induce such DAVN proofs of Bell nonlocality, but the construction for any of them is a big challenge. Moreover, note that if one use the extended constraints (See Tables \ref{TB1} and \ref{TB2}) in the argument of each HLQP paradox, one can find a homomorphism-like map between the algebraic relations leading to the contradiction and those in the qubit scenario (see a four-qubit example in Ref.\cite{Tang-DAVN-2022}). This may shed light on the further constructions of such proofs in the scenarios with $d=2^k$ ($k>1$).

\section{Conclusion}

To summarize, by using a combination technique, we have constructed a nontrivial DAVN proof of Bell nonlocality based on a four-qudit ($d=4$) non-stabilizer quantum state, filling the gap of the study of such proofs in qudit scenarios. Note that although we have used a special stabilizer ($Z_1Z_2\cdots Z_n$) to simplify the component chocie of the involved states, how to arrange the phase factor of each component still requires a tremendous  number of calculations (this is quite different from the qubit scenario, see Ref.\cite{Tang-DAVN-2022} ). This makes the construction of such a proof extremely challenging. So far, we have still not found a sufficiently efficient method (whatever by analytic construction or by computer search) to give general DAVN proofs of Bell nonlocality from qudit non-stabilizer states. In fact, even that whether one can construct such proofs from qudit  non-stabilizer states in the scenario of $d=3$ (or more generally, an odd number $d$) remains an open question.

on the other hand, our results can show the distinction of nonlocality and entanglement in some sense. To be specific, proofs of Bell nonlocality induced from the states with different entanglement may have the same  efficiency of ruling out the LHV model. In fact, one can find a four qudit ($d=4$) stabilizer state $|\Psi\rangle_{1234}^{\prime}$, and construct a GHZ-like proof of Bell nonlocality, which can rule out the LHV model with a probability of $100\%$, the same to the aforementioned example induced from the non-stabilizer state $|\Psi\rangle_{1234}$. By contrast, the entanglement of the stabilizer state $|\Psi\rangle_{1234}^{\prime}$ and the non-stabilizer state $|\Psi\rangle_{1234}$ is different.

Moreover, due to the difficulty of exploring such non-stabilizer quantum states (especially when the qudit number $n$ is large), sometimes even these states themselves can  be considered as a kind of rare resource (like bitcoins, large prime numbers). However, their applications  requires further study.

W. Du, D. Zhou and K. Han contribute equally to this work.

\acknowledgments
This work was supported by the National Nature Science Foundation of China (Grant No. 12271325).

\appendix
\renewcommand{\appendixname}{Appendix}
\section{The Hardy-like conditions for types IV, V, and VI}

Like the case of type III, we group the Hardy-like conditions in each type into two sub-classes, i.e., type IV-A, type IV-B;
type V-A, type V-B; type VI-A, type VI-B. They are listed in Tables \ref{TB5}-\ref{TB10} respectively.


\begin{table*}\caption{Six groups of Hardy-like conditions for type IV-A, in which the measurement results for $Z_1$, $Z_2$, $Z_3$, $Z_4$ arise from the contributions of the terms $i|0211\rangle$, $i|1021\rangle$, $i|1102\rangle$, $i|2110\rangle$, $i|0121\rangle$, $i|1012\rangle$, respectively.} \label{TB5}
    \centering
    \centering
    \begin{tabular}{c|ccc}
    \hline
    \hline
        ~~$Z_i,Z_j$ $(i\neq j)$~~ & ~~$|\psi\rangle_{kl}$ ($k\neq l\neq i\neq j$)~~ & ~~ Basic constraints  ~~  &  ~~ Extended constraints ~~\\
  \hline
 ~~$Z_1=1,Z_2=-1$~~  & ~~$|\psi\rangle_{34}=\frac{1}{2}(-|02\rangle+i|11\rangle-|20\rangle+i|33\rangle)$~~ &  ~~$X_3^2X_4^2=1$~~ & ~~$X_3^2X_4^2=1$~~\\
 ~~$Z_1=1,Z_3=i$~~  & ~~$|\psi\rangle_{24}=\frac{1}{2}(i|03\rangle-i|12\rangle+i|21\rangle-i|30\rangle)$~~ &  ~~$X_2X_4^3=-1$~~ & ~~$X_2^2X_4^2=1$~~ \\
 ~~$Z_1=1,Z_4=i$~~  & ~~$|\psi\rangle_{23}=\frac{1}{2}(-i|03\rangle+i|12\rangle+i|21\rangle-i|30\rangle)$~~ &  ~~$X_2^2X_3^2=-1$~~ & ~~$X_2^2X_3^2=-1$~~ \\
 ~~$Z_2=-1,Z_3=i$~~  & ~~$|\psi\rangle_{14}=\frac{1}{2}(i|01\rangle-i|10\rangle-i|23\rangle+i|32\rangle)$~~ &  ~~$X_1^2X_4^2=-1$~~ &  ~~$X_1^2X_4^2=-1$~~\\
 ~~$Z_2=-1,Z_4=i$~~  & ~~$|\psi\rangle_{13}=\frac{1}{2}(i|01\rangle-i|10\rangle+i|23\rangle-i|32\rangle)$~~ &  ~~$X_1X_3^3=-1$~~ & ~~$X_1^2X_3^2=1$~~ \\
 ~~$Z_3=i,Z_4=i$~~  & ~~$|\psi\rangle_{12}=\frac{1}{\sqrt{2}}(i|02\rangle-i|20\rangle)$~~ &  ~~$X_1^2X_2^2=-1$~~ & ~~$X_1^2X_2^2=-1$~~ \\
     \hline
 ~~$Z_1=i,Z_2=1$~~  & ~~$|\psi\rangle_{34}=\frac{1}{2}(-i|03\rangle+i|12\rangle+i|21\rangle+i|30\rangle)$~~ &  ~~$\text{---}$~~ & ~~$\text{---}$~~\\
 ~~$Z_1=i,Z_3=-1$~~  & ~~$|\psi\rangle_{24}=\frac{1}{2}(i|01\rangle-i|10\rangle+i|23\rangle-i|32\rangle)$~~ &  ~~$X_2X_4^3=-1$~~ & ~~$X_2^2X_4^2=1$~~ \\
 ~~$Z_1=i,Z_4=i$~~  & ~~$|\psi\rangle_{23}=\frac{1}{\sqrt{2}}(i|02\rangle-i|20\rangle)$~~ &  ~~$X_2^2X_3^2=-1$~~ & ~~$X_2^2X_3^2=-1$~~ \\
 ~~$Z_2=1,Z_3=-1$~~  & ~~$|\psi\rangle_{14}=\frac{1}{2}(-|02\rangle+i|11\rangle-|20\rangle+i|33\rangle)$~~ &  ~~$X_1^2X_4^2=1$~~ &  ~~$X_1^2X_4^2=1$~~\\
 ~~$Z_2=1,Z_4=i$~~  & ~~$|\psi\rangle_{13}=\frac{1}{2}(-i|03\rangle+i|12\rangle-i|21\rangle+i|30\rangle)$~~ &  ~~$X_1X_3^3=-1$~~ & ~~$X_1^2X_3^2=1$~~ \\
 ~~$Z_3=-1,Z_4=i$~~  & ~~$|\psi\rangle_{12}=\frac{1}{2}(i|01\rangle+i|10\rangle+i|23\rangle-i|32\rangle)$~~ &  ~~$\text{---}$~~ & ~~$\text{---}$~~ \\
     \hline
 ~~$Z_1=i,Z_2=i$~~  & ~~$|\psi\rangle_{34}=\frac{1}{\sqrt{2}}(i|02\rangle-i|20\rangle)$~~ &  ~~$X_3^2X_4^2=-1$~~ & ~~$X_3^2X_4^2=-1$~~\\
 ~~$Z_1=i,Z_3=1$~~  & ~~$|\psi\rangle_{24}=\frac{1}{2}(-i|03\rangle+i|12\rangle-i|21\rangle+i|30\rangle)$~~ &  ~~$X_2X_4^3=-1$~~ & ~~$X_2^2X_4^2=1$~~ \\
 ~~$Z_1=i,Z_4=-1$~~  & ~~$|\psi\rangle_{23}=\frac{1}{2}(i|01\rangle+i|10\rangle-i|23\rangle-i|32\rangle)$~~ &  ~~$X_2X_3^3=-1$~~ & ~~$X_2^2X_3^2=-1$~~ \\
 ~~$Z_2=i,Z_3=1$~~  & ~~$|\psi\rangle_{14}=\frac{1}{2}(i|03\rangle+i|12\rangle-i|21\rangle-i|30\rangle)$~~ &  ~~$X_1^2X_4^2=-1$~~ &  ~~$X_1^2X_4^2=-1$~~\\
 ~~$Z_2=i,Z_4=-1$~~  & ~~$|\psi\rangle_{13}=\frac{1}{2}(-i|01\rangle+i|10\rangle-i|23\rangle+i|32\rangle)$~~ &  ~~$X_1X_3^3=-1$~~ & ~~$X_1^2X_3^2=1$~~ \\
 ~~$Z_3=1,Z_4=-1$~~  & ~~$|\psi\rangle_{12}=\frac{1}{2}(-|02\rangle+i|11\rangle-|20\rangle+i|33\rangle)$~~ &  ~~$X_1^2X_2^2=1$~~ & ~~$X_1^2X_2^2=1$~~ \\
     \hline
 ~~$Z_1=-1,Z_2=i$~~  & ~~$|\psi\rangle_{34}=\frac{1}{2}(-i|01\rangle+i|10\rangle-i|23\rangle-i|32\rangle)$~~ &  ~~$\text{---}$~~ & ~~$\text{---}$~~\\
 ~~$Z_1=-1,Z_3=i$~~  & ~~$|\psi\rangle_{24}=\frac{1}{2}(-i|01\rangle+i|10\rangle-i|23\rangle+i|32\rangle)$~~ &  ~~$X_2X_4^3=-1$~~ & ~~$X_2^2X_4^2=1$~~ \\
 ~~$Z_1=-1,Z_4=1$~~  & ~~$|\psi\rangle_{23}=\frac{1}{2}(-|02\rangle+i|11\rangle-|20\rangle+i|33\rangle)$~~ &  ~~$X_2^2X_3^2=1$~~ & ~~$X_2^2X_3^2=1$~~ \\
 ~~$Z_2=i,Z_3=i$~~  & ~~$|\psi\rangle_{14}=\frac{1}{\sqrt{2}}(-i|02\rangle+i|20\rangle)$~~ &  ~~$X_1^2X_4^2=-1$~~ &  ~~$X_1^2X_4^2=-1$~~\\
 ~~$Z_2=i,Z_4=1$~~  & ~~$|\psi\rangle_{13}=\frac{1}{2}(i|03\rangle-i|12\rangle+i|21\rangle-i|30\rangle)$~~ &  ~~$X_1X_3^3=-1$~~ & ~~$X_1^2X_3^2=1$~~ \\
 ~~$Z_3=-i,Z_4=1$~~  & ~~$|\psi\rangle_{12}=\frac{1}{2}(-i|03\rangle-i|12\rangle+i|21\rangle-i|30\rangle)$~~ & ~~$\text{---}$~~ & ~~$\text{---}$~~\\
     \hline
 ~~$Z_1=1,Z_2=i$~~  & ~~$|\psi\rangle_{34}=\frac{1}{2}(i|03\rangle-i|12\rangle+i|21\rangle+i|30\rangle)$~~ &  ~~$\text{---}$~~ & ~~$\text{---}$~~\\
 ~~$Z_1=1,Z_3=-1$~~  & ~~$|\psi\rangle_{24}=\frac{1}{2}(-|02\rangle+i|11\rangle-|20\rangle+i|33\rangle)$~~ &  ~~$X_2^2X_4^2=1$~~ & ~~$X_2^2X_4^2=1$~~ \\
 ~~$Z_1=1,Z_4=i$~~  & ~~$|\psi\rangle_{23}=\frac{1}{2}(-i|03\rangle+i|12\rangle+i|21\rangle-i|30\rangle)$~~ &  ~~$X_2^2X_3^2=-1$~~ & ~~$X_2^2X_3^2=-1$~~ \\
 ~~$Z_2=-i,Z_3=-1$~~  & ~~$|\psi\rangle_{14}=\frac{1}{2}(i|01\rangle-i|10\rangle-i|23\rangle+i|32\rangle)$~~ &  ~~$X_1^2X_4^2=-1$~~ &  ~~$X_1^2X_4^2=-1$~~\\
 ~~$Z_2=i,Z_4=i$~~  & ~~$|\psi\rangle_{13}=\frac{1}{\sqrt{2}}(i|02\rangle-i|20\rangle)$~~ &  ~~$X_1^2X_3^2=-1$~~ & ~~$X_1^2X_3^2=-1$~~ \\
 ~~$Z_3=-1,Z_4=i$~~  & ~~$|\psi\rangle_{12}=\frac{1}{2}(i|01\rangle+i|10\rangle+i|23\rangle-i|32\rangle)$~~ &  ~~$\text{---}$~~ & ~~$\text{---}$~~ \\
     \hline
 ~~$Z_1=i,Z_2=1$~~  & ~~$|\psi\rangle_{34}=\frac{1}{2}(-i|03\rangle+i|12\rangle+i|21\rangle+i|30\rangle)$~~ & ~~$\text{---}$~~ & ~~$\text{---}$~~\\
 ~~$Z_1=i,Z_3=i$~~  & ~~$|\psi\rangle_{24}=\frac{1}{\sqrt{2}}(i|02\rangle-i|20\rangle)$~~ &  ~~$X_2^2X_4^2=-1$~~ & ~~$X_2^2X_4^2=-1$~~ \\
 ~~$Z_1=-i,Z_4=-1$~~  & ~~$|\psi\rangle_{23}=\frac{1}{2}(i|01\rangle+i|10\rangle-i|23\rangle-i|32\rangle)$~~ &  ~~$X_2^2X_3^2=-1$~~ & ~~$X_2^2X_3^2=-1$~~ \\
 ~~$Z_2=1,Z_3=i$~~  & ~~$|\psi\rangle_{14}=\frac{1}{2}(i|03\rangle+i|12\rangle-i|21\rangle-i|30\rangle)$~~ &  ~~$X_1^2X_4^2=-1$~~ &  ~~$X_1^2X_4^2=-1$~~\\
 ~~$Z_2=1,Z_4=-1$~~  & ~~$|\psi\rangle_{13}=\frac{1}{2}(-|02\rangle+i|11\rangle-|20\rangle+i|33\rangle)$~~ &  ~~$X_1^2X_3^2=1$~~ & ~~$X_1^2X_3^2=1$~~ \\
 ~~$Z_3=i,Z_4=-1$~~  & ~~$|\psi\rangle_{12}=\frac{1}{2}(-i|01\rangle+i|10\rangle+i|23\rangle+i|32\rangle)$~~ &  ~~$\text{---}$~~ & ~~$\text{---}$~~ \\
     \hline
 \hline
    \end{tabular}
\end{table*}

\begin{table*}\caption{Six groups of Hardy-like conditions for type IV-B, in which the measurement results for $Z_1$, $Z_2$, $Z_3$, $Z_4$ arise from the contributions of the terms $-i|2011\rangle$, $-i|1201\rangle$, $-i|1120\rangle$, $-i|0112\rangle$, $-i|2101\rangle$, $-i|1210\rangle$, respectively.} \label{TB6}
    \centering
    \centering
    \begin{tabular}{c|ccc}
    \hline
    \hline
        ~~$Z_i,Z_j$ $(i\neq j)$~~ & ~~$|\psi\rangle_{kl}$ ($k\neq l\neq i\neq j$)~~ & ~~ Basic constraints  ~~  &  ~~ Extended constraints ~~\\
  \hline
 ~~$Z_1=-1,Z_2=1$~~  & ~~$|\psi\rangle_{34}=\frac{1}{2}(-|02\rangle-i|11\rangle-|20\rangle-i|33\rangle)$~~ &  ~~$X_3^2X_4^2=1$~~ & ~~$X_3^2X_4^2=1$~~\\
 ~~$Z_1=-1,Z_3=i$~~  & ~~$|\psi\rangle_{24}=\frac{1}{2}(-i|01\rangle+i|10\rangle-i|23\rangle+i|32\rangle)$~~ &  ~~$X_2X_4^3=-1$~~ & ~~$X_2^2X_4^2=1$~~ \\
 ~~$Z_1=-1,Z_4=i$~~  & ~~$|\psi\rangle_{23}=\frac{1}{2}(-i|01\rangle-i|10\rangle+i|23\rangle+i|32\rangle)$~~ &  ~~$X_2^2X_3^2=-1$~~ & ~~$X_2^2X_3^2=-1$~~ \\
 ~~$Z_2=1,Z_3=i$~~  & ~~$|\psi\rangle_{14}=\frac{1}{2}(i|03\rangle+i|12\rangle-i|21\rangle-i|30\rangle)$~~ &  ~~$X_1^2X_4^2=-1$~~ &  ~~$X_1^2X_4^2=-1$~~\\
 ~~$Z_2=1,Z_4=i$~~  & ~~$|\psi\rangle_{13}=\frac{1}{2}(-i|03\rangle+i|12\rangle-i|21\rangle+i|30\rangle)$~~ &  ~~$X_1X_3^3=-1$~~ & ~~$X_1^2X_3^2=1$~~ \\
 ~~$Z_3=i,Z_4=i$~~  & ~~$|\psi\rangle_{12}=\frac{1}{\sqrt{2}}(i|02\rangle-i|20\rangle)$~~ &  ~~$X_1^2X_2^2=-1$~~ & ~~$X_1^2X_2^2=-1$~~ \\
     \hline
 ~~$Z_1=i,Z_2=-1$~~  & ~~$|\psi\rangle_{34}=\frac{1}{2}(-i|01\rangle-i|10\rangle+i|23\rangle-i|32\rangle)$~~ &  ~~$\text{---}$~~ & ~~$\text{---}$~~\\
 ~~$Z_1=i,Z_3=1$~~  & ~~$|\psi\rangle_{24}=\frac{1}{2}(-i|03\rangle+i|12\rangle-i|21\rangle+i|30\rangle)$~~ &  ~~$X_2X_4^3=-1$~~ & ~~$X_2^2X_4^2=1$~~ \\
 ~~$Z_1=i,Z_4=i$~~  & ~~$|\psi\rangle_{23}=\frac{1}{\sqrt{2}}(i|02\rangle-i|20\rangle)$~~ &  ~~$X_2^2X_3^2=-1$~~ & ~~$X_2^2X_3^2=-1$~~ \\
 ~~$Z_2=-1,Z_3=1$~~  & ~~$|\psi\rangle_{14}=\frac{1}{2}(-|02\rangle-i|11\rangle-|20\rangle-i|33\rangle)$~~ &  ~~$X_1^2X_4^2=1$~~ &  ~~$X_1^2X_4^2=1$~~\\
 ~~$Z_2=-1,Z_4=i$~~  & ~~$|\psi\rangle_{13}=\frac{1}{2}(i|01\rangle-i|10\rangle+i|23\rangle-i|32\rangle)$~~ &  ~~$X_1X_3^3=-1$~~ & ~~$X_1^2X_3^2=1$~~ \\
 ~~$Z_3=1,Z_4=i$~~  & ~~$|\psi\rangle_{12}=\frac{1}{2}(-i|03\rangle-i|12\rangle-i|21\rangle+i|30\rangle)$~~ &  ~~$\text{---}$~~ & ~~$\text{---}$~~ \\
     \hline
 ~~$Z_1=i,Z_2=i$~~  & ~~$|\psi\rangle_{34}=\frac{1}{\sqrt{2}}(i|02\rangle-i|20\rangle)$~~ &  ~~$X_3^2X_4^2=-1$~~ & ~~$X_3^2X_4^2=-1$~~\\
 ~~$Z_1=i,Z_3=-1$~~  & ~~$|\psi\rangle_{24}=\frac{1}{2}(i|01\rangle-i|10\rangle+i|23\rangle-i|32\rangle)$~~ &  ~~$X_2X_4^3=-1$~~ & ~~$X_2^2X_4^2=1$~~ \\
 ~~$Z_1=i,Z_4=1$~~  & ~~$|\psi\rangle_{23}=\frac{1}{2}(i|03\rangle-i|12\rangle-i|21\rangle+i|30\rangle)$~~ &  ~~$X_2^2X_3^2=-1$~~ & ~~$X_2^2X_3^2=-1$~~ \\
 ~~$Z_2=i,Z_3=-1$~~  & ~~$|\psi\rangle_{14}=\frac{1}{2}(i|01\rangle-i|10\rangle-i|23\rangle+i|32\rangle)$~~ &  ~~$X_1^2X_4^2=-1$~~ &  ~~$X_1^2X_4^2=-1$~~\\
 ~~$Z_2=i,Z_4=1$~~  & ~~$|\psi\rangle_{13}=\frac{1}{2}(i|03\rangle-i|12\rangle+i|21\rangle-i|30\rangle)$~~ &  ~~$X_1X_3^3=-1$~~ & ~~$X_1^2X_3^2=1$~~ \\
 ~~$Z_3=-1,Z_4=1$~~  & ~~$|\psi\rangle_{12}=\frac{1}{2}(-|02\rangle-i|11\rangle-|20\rangle-i|33\rangle)$~~ &  ~~$X_1^2X_2^2=1$~~ & ~~$X_1^2X_2^2=1$~~ \\
     \hline
 ~~$Z_1=1,Z_2=i$~~  & ~~$|\psi\rangle_{34}=\frac{1}{2}(i|03\rangle-i|12\rangle+i|21\rangle+i|30\rangle)$~~ &  ~~$\text{---}$~~ & ~~$\text{---}$~~\\
 ~~$Z_1=1,Z_3=i$~~  & ~~$|\psi\rangle_{24}=\frac{1}{2}(i|03\rangle-i|12\rangle+i|21\rangle-i|30\rangle)$~~ &  ~~$X_2X_4^3=-1$~~ & ~~$X_2^2X_4^2=1$~~ \\
 ~~$Z_1=1,Z_4=-1$~~  & ~~$|\psi\rangle_{23}=\frac{1}{2}(-|02\rangle-i|11\rangle-|20\rangle-i|33\rangle)$~~ &  ~~$X_2^2X_3^2=1$~~ & ~~$X_2^2X_3^2=1$~~ \\
 ~~$Z_2=i,Z_3=i$~~  & ~~$|\psi\rangle_{14}=\frac{1}{\sqrt{2}}(-i|02\rangle+i|20\rangle)$~~ &  ~~$X_1^2X_4^2=-1$~~ &  ~~$X_1^2X_4^2=-1$~~\\
 ~~$Z_2=i,Z_4=-1$~~  & ~~$|\psi\rangle_{13}=\frac{1}{2}(-i|01\rangle+i|10\rangle-i|23\rangle+i|32\rangle)$~~ &  ~~$X_1X_3^3=-1$~~ & ~~$X_1^2X_3^2=1$~~ \\
 ~~$Z_3=i,Z_4=-1$~~  & ~~$|\psi\rangle_{12}=\frac{1}{2}(-i|01\rangle+i|10\rangle+i|23\rangle+i|32\rangle)$~~ & ~~$\text{---}$~~ & ~~$\text{---}$~~\\
     \hline
 ~~$Z_1=-1,Z_2=i$~~  & ~~$|\psi\rangle_{34}=\frac{1}{2}(-i|01\rangle+i|10\rangle-i|23\rangle-i|32\rangle)$~~ &  ~~$\text{---}$~~ & ~~$\text{---}$~~\\
 ~~$Z_1=-1,Z_3=1$~~  & ~~$|\psi\rangle_{24}=\frac{1}{2}(-|02\rangle-i|11\rangle-|20\rangle-i|33\rangle)$~~ &  ~~$X_2^2X_4^2=1$~~ & ~~$X_2^2X_4^2=1$~~ \\
 ~~$Z_1=-1,Z_4=i$~~  & ~~$|\psi\rangle_{23}=\frac{1}{2}(-i|01\rangle-i|10\rangle+i|23\rangle+i|32\rangle)$~~ &  ~~$X_2^2X_3^2=-1$~~ & ~~$X_2^2X_3^2=-1$~~ \\
 ~~$Z_2=i,Z_3=1$~~  & ~~$|\psi\rangle_{14}=\frac{1}{2}(i|03\rangle+i|12\rangle-i|21\rangle-i|30\rangle)$~~ &  ~~$X_1^2X_4^2=-1$~~ &  ~~$X_1^2X_4^2=-1$~~\\
 ~~$Z_2=i,Z_4=i$~~  & ~~$|\psi\rangle_{13}=\frac{1}{\sqrt{2}}(i|02\rangle-i|20\rangle)$~~ &  ~~$X_1^2X_3^2=-1$~~ & ~~$X_1^2X_3^2=-1$~~ \\
 ~~$Z_3=1,Z_4=i$~~  & ~~$|\psi\rangle_{12}=\frac{1}{2}(-i|03\rangle-i|12\rangle-i|21\rangle+i|30\rangle)$~~ &  ~~$\text{---}$~~ & ~~$\text{---}$~~ \\
     \hline
 ~~$Z_1=i,Z_2=-1$~~  & ~~$|\psi\rangle_{34}=\frac{1}{2}(-i|01\rangle-i|10\rangle+i|23\rangle-i|32\rangle)$~~ & ~~$\text{---}$~~ & ~~$\text{---}$~~\\
 ~~$Z_1=i,Z_3=i$~~  & ~~$|\psi\rangle_{24}=\frac{1}{\sqrt{2}}(i|02\rangle-i|20\rangle)$~~ &  ~~$X_2^2X_4^2=-1$~~ & ~~$X_2^2X_4^2=-1$~~ \\
 ~~$Z_1=i,Z_4=1$~~  & ~~$|\psi\rangle_{23}=\frac{1}{2}(i|03\rangle-i|12\rangle-i|21\rangle+i|30\rangle)$~~ &  ~~$X_2^2X_3^2=-1$~~ & ~~$X_2^2X_3^2=-1$~~ \\
 ~~$Z_2=-1,Z_3=i$~~  & ~~$|\psi\rangle_{14}=\frac{1}{2}(i|01\rangle-i|10\rangle-i|23\rangle+i|32\rangle)$~~ &  ~~$X_1^2X_4^2=-1$~~ &  ~~$X_1^2X_4^2=-1$~~\\
 ~~$Z_2=-1,Z_4=1$~~  & ~~$|\psi\rangle_{13}=\frac{1}{2}(-|02\rangle-i|11\rangle-|20\rangle-i|33\rangle)$~~ &  ~~$X_1^2X_3^2=1$~~ & ~~$X_1^2X_3^2=1$~~ \\
 ~~$Z_3=i,Z_4=1$~~  & ~~$|\psi\rangle_{12}=\frac{1}{2}(-i|03\rangle-i|12\rangle+i|21\rangle-i|30\rangle)$~~ &  ~~$\text{---}$~~ & ~~$\text{---}$~~ \\
     \hline
 \hline
    \end{tabular}
\end{table*}


\begin{table*}\caption{Six groups of Hardy-like conditions for type V-A, in which the measurement results for $Z_1$, $Z_2$, $Z_3$, $Z_4$ arise from the contributions of the terms $i|1300\rangle$, $i|0130\rangle$, $i|0013\rangle$, $i|3001\rangle$, $i|1030\rangle$, $i|0103\rangle$, respectively.} \label{TB7}
    \centering
    \centering
    \begin{tabular}{c|ccc}
    \hline
    \hline
        ~~$Z_i,Z_j$ $(i\neq j)$~~ & ~~$|\psi\rangle_{kl}$ ($k\neq l\neq i\neq j$)~~ & ~~ Basic constraints  ~~  &  ~~ Extended constraints ~~\\
  \hline
 ~~$Z_1=i,Z_2=-i$~~  & ~~$|\psi\rangle_{34}=\frac{1}{\sqrt{2}}(i|00\rangle-i|22\rangle)$~~ &  ~~$X_3^2X_4^2=-1$~~ & ~~$X_3^2X_4^2=-1$~~\\
 ~~$Z_1=i,Z_3=1$~~  & ~~$|\psi\rangle_{24}=\frac{1}{2}(-i|03\rangle+i|12\rangle-i|21\rangle+i|30\rangle)$~~ &  ~~$X_2X_4^3=-1$~~ & ~~$X_2^2X_4^2=1$~~ \\
 ~~$Z_1=i,Z_4=1$~~  & ~~$|\psi\rangle_{23}=\frac{1}{2}(i|03\rangle-i|12\rangle-i|21\rangle+i|30\rangle)$~~ &  ~~$X_2^2X_3^2=-1$~~ & ~~$X_2^2X_3^2=-1$~~ \\
 ~~$Z_2=-i,Z_3=1$~~  & ~~$|\psi\rangle_{14}=\frac{1}{2}(-i|01\rangle+i|10\rangle-i|23\rangle+i|32\rangle)$~~ &  ~~$X_1X_4^3=-1$~~ &  ~~$X_1^2X_4^2=1$~~\\
 ~~$Z_2=-i,Z_4=1$~~  & ~~$|\psi\rangle_{13}=\frac{1}{2}(-i|01\rangle+i|10\rangle+i|23\rangle-i|32\rangle)$~~ &  ~~$X_1^2X_3^2=-1$~~ & ~~$X_1^2X_3^2=-1$~~ \\
 ~~$Z_3=1,Z_4=1$~~  & ~~$|\psi\rangle_{12}=\frac{1}{2}(|00\rangle+i|13\rangle-|22\rangle-i|31\rangle)$~~ &  ~~$X_1X_2^3=-i$~~ & ~~$X_1^2X_2^2=-1$~~ \\
     \hline
 ~~$Z_1=1,Z_2=i$~~  & ~~$|\psi\rangle_{34}=\frac{1}{2}(i|03\rangle-i|12\rangle+i|21\rangle+i|30\rangle)$~~ &  ~~$\text{---}$~~ & ~~$\text{---}$~~\\
 ~~$Z_1=1,Z_3=-i$~~  & ~~$|\psi\rangle_{24}=\frac{1}{2}(-i|01\rangle+i|10\rangle+i|23\rangle-i|32\rangle)$~~ &  ~~$X_2^2X_4^2=-1$~~ & ~~$X_2^2X_4^2=-1$~~ \\
 ~~$Z_1=1,Z_4=1$~~  & ~~$|\psi\rangle_{23}=\frac{1}{2}(|00\rangle+i|13\rangle-|22\rangle-i|31\rangle)$~~ &  ~~$X_2X_3^3=-i$~~ & ~~$X_2^2X_3^2=-1$~~ \\
 ~~$Z_2=i,Z_3=-i$~~  & ~~$|\psi\rangle_{14}=\frac{1}{\sqrt{2}}(i|00\rangle-i|22\rangle)$~~ &  ~~$X_1^2X_4^2=-1$~~ &  ~~$X_1^2X_4^2=-1$~~\\
 ~~$Z_2=i,Z_4=1$~~  & ~~$|\psi\rangle_{13}=\frac{1}{2}(i|03\rangle-i|12\rangle+i|21\rangle-i|30\rangle)$~~ &  ~~$X_1X_3^3=-1$~~ & ~~$X_1^2X_3^2=1$~~ \\
 ~~$Z_3=-i,Z_4=1$~~  & ~~$|\psi\rangle_{12}=\frac{1}{2}(i|01\rangle+i|10\rangle+i|23\rangle-i|32\rangle)$~~ &  ~~$\text{---}$~~ & ~~$\text{---}$~~ \\
     \hline
 ~~$Z_1=1,Z_2=1$~~  & ~~$|\psi\rangle_{34}=\frac{1}{2}(|00\rangle+i|13\rangle-|22\rangle-i|31\rangle)$~~ &  ~~$X_3X_4^3=-i$~~ & ~~$X_3^2X_4^2=-1$~~\\
 ~~$Z_1=1,Z_3=i$~~  & ~~$|\psi\rangle_{24}=\frac{1}{2}(i|03\rangle-i|12\rangle+i|21\rangle-i|30\rangle)$~~ &  ~~$X_2X_4^3=-1$~~ & ~~$X_2^2X_4^2=1$~~ \\
 ~~$Z_1=1,Z_4=-i$~~  & ~~$|\psi\rangle_{23}=\frac{1}{2}(i|01\rangle+i|10\rangle+i|23\rangle+i|32\rangle)$~~ &  ~~$X_2X_3^3=1$~~ & ~~$X_2^2X_3^2=1$~~ \\
 ~~$Z_2=1,Z_3=i$~~  & ~~$|\psi\rangle_{14}=\frac{1}{2}(i|03\rangle+i|12\rangle-i|21\rangle-i|30\rangle)$~~ &  ~~$X_1^2X_4^2=-1$~~ &  ~~$X_1^2X_4^2=-1$~~\\
 ~~$Z_2=1,Z_4=-i$~~  & ~~$|\psi\rangle_{13}=\frac{1}{2}(i|01\rangle-i|10\rangle-i|23\rangle+i|32\rangle)$~~ &  ~~$X_1^2X_3^2=-1$~~ & ~~$X_1^2X_3^2=-1$~~ \\
 ~~$Z_3=i,Z_4=-i$~~  & ~~$|\psi\rangle_{12}=\frac{1}{\sqrt{2}}(i|00\rangle-i|22\rangle)$~~ &  ~~$X_1^2X_2^2=-1$~~ & ~~$X_1^2X_2^2=-1$~~ \\
     \hline
 ~~$Z_1=-i,Z_2=1$~~  & ~~$|\psi\rangle_{34}=\frac{1}{2}(i|01\rangle-i|10\rangle+i|23\rangle+i|32\rangle)$~~ &  ~~$\text{---}$~~ & ~~$\text{---}$~~\\
 ~~$Z_1=-i,Z_3=1$~~  & ~~$|\psi\rangle_{24}=\frac{1}{2}(i|01\rangle-i|10\rangle-i|23\rangle+i|32\rangle)$~~ &  ~~$X_2^2X_4^2=-1$~~ & ~~$X_2^2X_4^2=-1$~~ \\
 ~~$Z_1=-i,Z_4=i$~~  & ~~$|\psi\rangle_{23}=\frac{1}{\sqrt{2}}(i|00\rangle-i|22\rangle)$~~ &  ~~$X_2^2X_3^2=-1$~~ & ~~$X_2^2X_3^2=-1$~~ \\
 ~~$Z_2=1,Z_3=1$~~  & ~~$|\psi\rangle_{14}=\frac{1}{2}(|00\rangle-i|13\rangle-|22\rangle+i|31\rangle)$~~ &  ~~$X_1X_4^3=i$~~ &  ~~$X_1^2X_4^2=-1$~~\\
 ~~$Z_2=1,Z_4=i$~~  & ~~$|\psi\rangle_{13}=\frac{1}{2}(-i|03\rangle+i|12\rangle-i|21\rangle+i|30\rangle)$~~ &  ~~$X_1X_3^3=-1$~~ & ~~$X_1^2X_3^2=1$~~ \\
 ~~$Z_3=1,Z_4=i$~~  & ~~$|\psi\rangle_{12}=\frac{1}{2}(-i|03\rangle-i|12\rangle-i|21\rangle+i|30\rangle)$~~ & ~~$\text{---}$~~ & ~~$\text{---}$~~\\
     \hline
 ~~$Z_1=i,Z_2=1$~~  & ~~$|\psi\rangle_{34}=\frac{1}{2}(-i|03\rangle+i|12\rangle+i|21\rangle+i|30\rangle)$~~ &  ~~$\text{---}$~~ & ~~$\text{---}$~~\\
 ~~$Z_1=i,Z_3=-i$~~  & ~~$|\psi\rangle_{24}=\frac{1}{\sqrt{2}}(i|00\rangle-i|22\rangle)$~~ &  ~~$X_2^2X_4^2=-1$~~ & ~~$X_2^2X_4^2=-1$~~ \\
 ~~$Z_1=i,Z_4=1$~~  & ~~$|\psi\rangle_{23}=\frac{1}{2}(i|03\rangle-i|12\rangle-i|21\rangle+i|30\rangle)$~~ &  ~~$X_2^2X_3^2=-1$~~ & ~~$X_2^2X_3^2=-1$~~ \\
 ~~$Z_2=1,Z_3=-i$~~  & ~~$|\psi\rangle_{14}=\frac{1}{2}(-i|01\rangle+i|10\rangle-i|23\rangle+i|32\rangle)$~~ &  ~~$X_1X_4^3=-1$~~ &  ~~$X_1^2X_4^2=1$~~\\
 ~~$Z_2=1,Z_4=1$~~  & ~~$|\psi\rangle_{13}=\frac{1}{2}(|00\rangle+i|13\rangle-|22\rangle-i|31\rangle)$~~ &  ~~$X_1X_3^3=-i$~~ & ~~$X_1^2X_3^2=-1$~~ \\
 ~~$Z_3=-i,Z_4=1$~~  & ~~$|\psi\rangle_{12}=\frac{1}{2}(i|01\rangle+i|10\rangle+i|23\rangle-i|32\rangle)$~~ &  ~~$\text{---}$~~ & ~~$\text{---}$~~ \\
     \hline
 ~~$Z_1=1,Z_2=i$~~  & ~~$|\psi\rangle_{34}=\frac{1}{2}(i|03\rangle-i|12\rangle+i|21\rangle+i|30\rangle)$~~ & ~~$\text{---}$~~ & ~~$\text{---}$~~\\
 ~~$Z_1=1,Z_3=1$~~  & ~~$|\psi\rangle_{24}=\frac{1}{2}(|00\rangle+i|13\rangle-|22\rangle-i|31\rangle)$~~ &  ~~$X_2X_4^3=-i$~~ & ~~$X_2^2X_4^2=-1$~~ \\
 ~~$Z_1=1,Z_4=-i$~~  & ~~$|\psi\rangle_{23}=\frac{1}{2}(i|01\rangle+i|10\rangle+i|23\rangle+i|32\rangle)$~~ &  ~~$X_2X_3^3=1$~~ & ~~$X_2^2X_3^2=1$~~ \\
 ~~$Z_2=i,Z_3=1$~~  & ~~$|\psi\rangle_{14}=\frac{1}{2}(i|03\rangle+i|12\rangle-i|21\rangle-i|30\rangle)$~~ &  ~~$X_1^2X_4^2=-1$~~ &  ~~$X_1^2X_4^2=-1$~~\\
 ~~$Z_2=i,Z_4=-i$~~  & ~~$|\psi\rangle_{13}=\frac{1}{\sqrt{2}}(i|00\rangle-i|22\rangle)$~~ &  ~~$X_1^2X_3^2=-1$~~ & ~~$X_1^2X_3^2=-1$~~ \\
 ~~$Z_3=1,Z_4=-i$~~  & ~~$|\psi\rangle_{12}=\frac{1}{2}(i|01\rangle-i|10\rangle-i|23\rangle-i|32\rangle)$~~ &  ~~$\text{---}$~~ & ~~$\text{---}$~~ \\
     \hline
 \hline
    \end{tabular}
\end{table*}

\begin{table*}\caption{Six groups of Hardy-like conditions for type V-B, in which the measurement results for $Z_1$, $Z_2$, $Z_3$, $Z_4$ arise from the contributions of the terms $-i|3100\rangle$, $-i|0310\rangle$, $-i|0031\rangle$, $-i|1003\rangle$, $-i|3010\rangle$, $-i|0301\rangle$, respectively.} \label{TB8}
    \centering
    \centering
    \begin{tabular}{c|ccc}
    \hline
    \hline
        ~~$Z_i,Z_j$ $(i\neq j)$~~ & ~~$|\psi\rangle_{kl}$ ($k\neq l\neq i\neq j$)~~ & ~~ Basic constraints  ~~  &  ~~ Extended constraints ~~\\
  \hline
 ~~$Z_1=-i,Z_2=i$~~  & ~~$|\psi\rangle_{34}=\frac{1}{\sqrt{2}}(-i|00\rangle+i|22\rangle)$~~ &  ~~$X_3^2X_4^2=-1$~~ & ~~$X_3^2X_4^2=-1$~~\\
 ~~$Z_1=-i,Z_3=1$~~  & ~~$|\psi\rangle_{24}=\frac{1}{2}(i|01\rangle-i|10\rangle-i|23\rangle+i|32\rangle)$~~ &  ~~$X_2^2X_4^2=-1$~~ & ~~$X_2^2X_4^2=-1$~~ \\
 ~~$Z_1=-i,Z_4=1$~~  & ~~$|\psi\rangle_{23}=\frac{1}{2}(-i|01\rangle-i|10\rangle-i|23\rangle-i|32\rangle)$~~ &  ~~$X_2X_3^3=1$~~ & ~~$X_2^2X_3^2=1$~~ \\
 ~~$Z_2=i,Z_3=1$~~  & ~~$|\psi\rangle_{14}=\frac{1}{2}(i|03\rangle+i|12\rangle-i|21\rangle-i|30\rangle)$~~ &  ~~$X_1^2X_4^2=-1$~~ &  ~~$X_1^2X_4^2=-1$~~\\
 ~~$Z_2=i,Z_4=1$~~  & ~~$|\psi\rangle_{13}=\frac{1}{2}(i|03\rangle-i|12\rangle+i|21\rangle-i|30\rangle)$~~ &  ~~$X_1X_3^3=-1$~~ & ~~$X_1^2X_3^2=1$~~ \\
 ~~$Z_3=1,Z_4=1$~~  & ~~$|\psi\rangle_{12}=\frac{1}{2}(|00\rangle+i|13\rangle-|22\rangle-i|31\rangle)$~~ &  ~~$X_1X_2^3=-i$~~ & ~~$X_1^2X_2^2=-1$~~ \\
     \hline
 ~~$Z_1=1,Z_2=-i$~~  & ~~$|\psi\rangle_{34}=\frac{1}{2}(-i|01\rangle-i|10\rangle+i|23\rangle-i|32\rangle)$~~ &  ~~$\text{---}$~~ & ~~$\text{---}$~~\\
 ~~$Z_1=1,Z_3=i$~~  & ~~$|\psi\rangle_{24}=\frac{1}{2}(i|03\rangle-i|12\rangle+i|21\rangle-i|30\rangle)$~~ &  ~~$X_2X_4^3=-1$~~ & ~~$X_2^2X_4^2=1$~~ \\
 ~~$Z_1=1,Z_4=1$~~  & ~~$|\psi\rangle_{23}=\frac{1}{2}(|00\rangle+i|13\rangle-|22\rangle-i|31\rangle)$~~ &  ~~$X_2X_3^3=-i$~~ & ~~$X_2^2X_3^2=-1$~~ \\
 ~~$Z_2=-i,Z_3=i$~~  & ~~$|\psi\rangle_{14}=\frac{1}{\sqrt{2}}(-i|00\rangle+i|22\rangle)$~~ &  ~~$X_1^2X_4^2=-1$~~ &  ~~$X_1^2X_4^2=-1$~~\\
 ~~$Z_2=-i,Z_4=1$~~  & ~~$|\psi\rangle_{13}=\frac{1}{2}(-i|01\rangle+i|10\rangle+i|23\rangle-i|32\rangle)$~~ &  ~~$X_1^2X_3^2=-1$~~ & ~~$X_1^2X_3^2=-1$~~ \\
 ~~$Z_3=i,Z_4=1$~~  & ~~$|\psi\rangle_{12}=\frac{1}{2}(-i|03\rangle-i|12\rangle+i|21\rangle-i|30\rangle)$~~ &  ~~$\text{---}$~~ & ~~$\text{---}$~~ \\
     \hline
 ~~$Z_1=1,Z_2=1$~~  & ~~$|\psi\rangle_{34}=\frac{1}{2}(|00\rangle+i|13\rangle-|22\rangle-i|31\rangle)$~~ &  ~~$X_3X_4^3=-i$~~ & ~~$X_3^2X_4^2=-1$~~\\
 ~~$Z_1=1,Z_3=-i$~~  & ~~$|\psi\rangle_{24}=\frac{1}{2}(-i|01\rangle+i|10\rangle+i|23\rangle-i|32\rangle)$~~ &  ~~$X_2^2X_4^2=-1$~~ & ~~$X_2^2X_4^2=-1$~~ \\
 ~~$Z_1=1,Z_4=i$~~  & ~~$|\psi\rangle_{23}=\frac{1}{2}(-i|03\rangle+i|12\rangle+i|21\rangle-i|30\rangle)$~~ &  ~~$X_2^2X_3^2=-1$~~ & ~~$X_2^2X_3^2=-1$~~ \\
 ~~$Z_2=1,Z_3=-i$~~  & ~~$|\psi\rangle_{14}=\frac{1}{2}(-i|01\rangle+i|10\rangle-i|23\rangle+i|32\rangle)$~~ &  ~~$X_1X_4^3=-1$~~ &  ~~$X_1^2X_4^2=1$~~\\
 ~~$Z_2=1,Z_4=i$~~  & ~~$|\psi\rangle_{13}=\frac{1}{2}(-i|03\rangle+i|12\rangle-i|21\rangle+i|30\rangle)$~~ &  ~~$X_1X_3^3=-1$~~ & ~~$X_1^2X_3^2=1$~~ \\
 ~~$Z_3=-i,Z_4=i$~~  & ~~$|\psi\rangle_{12}=\frac{1}{\sqrt{2}}(-i|00\rangle+i|22\rangle)$~~ &  ~~$X_1^2X_2^2=-1$~~ & ~~$X_1^2X_2^2=-1$~~ \\
     \hline
 ~~$Z_1=i,Z_2=1$~~  & ~~$|\psi\rangle_{34}=\frac{1}{2}(-i|03\rangle+i|12\rangle+i|21\rangle+i|30\rangle)$~~ &  ~~$\text{---}$~~ & ~~$\text{---}$~~\\
 ~~$Z_1=i,Z_3=1$~~  & ~~$|\psi\rangle_{24}=\frac{1}{2}(-i|03\rangle+i|12\rangle-i|21\rangle+i|30\rangle)$~~ &  ~~$X_2X_4^3=-1$~~ & ~~$X_2^2X_4^2=1$~~ \\
 ~~$Z_1=i,Z_4=-i$~~  & ~~$|\psi\rangle_{23}=\frac{1}{\sqrt{2}}(-i|00\rangle+i|22\rangle)$~~ &  ~~$X_2^2X_3^2=-1$~~ & ~~$X_2^2X_3^2=-1$~~ \\
 ~~$Z_2=1,Z_3=1$~~  & ~~$|\psi\rangle_{14}=\frac{1}{2}(|00\rangle-i|13\rangle-|22\rangle+i|31\rangle)$~~ &  ~~$X_1X_4^3=i$~~ &  ~~$X_1^2X_4^2=-1$~~\\
 ~~$Z_2=1,Z_4=-i$~~  & ~~$|\psi\rangle_{13}=\frac{1}{2}(i|01\rangle-i|10\rangle-i|23\rangle+i|32\rangle)$~~ &  ~~$X_1^2X_3^2=-1$~~ & ~~$X_1^2X_3^2=-1$~~ \\
 ~~$Z_3=1,Z_4=-i$~~  & ~~$|\psi\rangle_{12}=\frac{1}{2}(i|01\rangle-i|10\rangle-i|23\rangle-i|32\rangle)$~~ & ~~$\text{---}$~~ & ~~$\text{---}$~~\\
     \hline
 ~~$Z_1=-i,Z_2=1$~~  & ~~$|\psi\rangle_{34}=\frac{1}{2}(i|01\rangle-i|10\rangle+i|23\rangle+i|32\rangle)$~~ &  ~~$\text{---}$~~ & ~~$\text{---}$~~\\
 ~~$Z_1=-i,Z_3=i$~~  & ~~$|\psi\rangle_{24}=\frac{1}{\sqrt{2}}(-i|00\rangle+i|22\rangle)$~~ &  ~~$X_2^2X_4^2=-1$~~ & ~~$X_2^2X_4^2=-1$~~ \\
 ~~$Z_1=-i,Z_4=1$~~  & ~~$|\psi\rangle_{23}=\frac{1}{2}(-i|01\rangle-i|10\rangle-i|23\rangle-i|32\rangle)$~~ &  ~~$X_2X_3^3=1$~~ & ~~$X_2^2X_3^2=1$~~ \\
 ~~$Z_2=1,Z_3=i$~~  & ~~$|\psi\rangle_{14}=\frac{1}{2}(i|03\rangle+i|12\rangle-i|21\rangle-i|30\rangle)$~~ &  ~~$X_1^2X_4^2=-1$~~ &  ~~$X_1^2X_4^2=-1$~~\\
 ~~$Z_2=1,Z_4=1$~~  & ~~$|\psi\rangle_{13}=\frac{1}{2}(|00\rangle+i|13\rangle-|22\rangle-i|31\rangle)$~~ &  ~~$X_1X_3^3=-i$~~ & ~~$X_1^2X_3^2=-1$~~ \\
 ~~$Z_3=i,Z_4=1$~~  & ~~$|\psi\rangle_{12}=\frac{1}{2}(-i|03\rangle-i|12\rangle+i|21\rangle-i|30\rangle)$~~ &  ~~$\text{---}$~~ & ~~$\text{---}$~~ \\
     \hline
 ~~$Z_1=1,Z_2=-i$~~  & ~~$|\psi\rangle_{34}=\frac{1}{2}(-i|01\rangle-i|10\rangle+i|23\rangle-i|32\rangle)$~~ & ~~$\text{---}$~~ & ~~$\text{---}$~~\\
 ~~$Z_1=1,Z_3=1$~~  & ~~$|\psi\rangle_{24}=\frac{1}{2}(|00\rangle+i|13\rangle-|22\rangle-i|31\rangle)$~~ &  ~~$X_2X_4^3=-i$~~ & ~~$X_2^2X_4^2=-1$~~ \\
 ~~$Z_1=1,Z_4=i$~~  & ~~$|\psi\rangle_{23}=\frac{1}{2}(-i|03\rangle+i|12\rangle+i|21\rangle-i|30\rangle)$~~ &  ~~$X_2^2X_3^2=-1$~~ & ~~$X_2^2X_3^2=-1$~~ \\
 ~~$Z_2=-i,Z_3=1$~~  & ~~$|\psi\rangle_{14}=\frac{1}{2}(-i|01\rangle+i|10\rangle-i|23\rangle+i|32\rangle)$~~ &  ~~$X_1X_4^3=-1$~~ &  ~~$X_1^2X_4^2=1$~~\\
 ~~$Z_2=-i,Z_4=i$~~  & ~~$|\psi\rangle_{13}=\frac{1}{\sqrt{2}}(-i|00\rangle+i|22\rangle)$~~ &  ~~$X_1^2X_3^2=-1$~~ & ~~$X_1^2X_3^2=-1$~~ \\
 ~~$Z_3=1,Z_4=i$~~  & ~~$|\psi\rangle_{12}=\frac{1}{2}(-i|03\rangle-i|12\rangle-i|21\rangle+i|30\rangle)$~~ &  ~~$\text{---}$~~ & ~~$\text{---}$~~ \\
     \hline
 \hline
    \end{tabular}
\end{table*}


\begin{table*}\caption{Six groups of Hardy-like conditions for type VI-A, in which the measurement results for $Z_1$, $Z_2$, $Z_3$, $Z_4$ arise from the contributions of the terms $-i|1322\rangle$, $-i|2132\rangle$, $-i|2213\rangle$, $-i|3221\rangle$, $-i|1232\rangle$, $-i|2123\rangle$, respectively.} \label{TB9}
    \centering
    \centering
    \begin{tabular}{c|ccc}
    \hline
    \hline
        ~~$Z_i,Z_j$ $(i\neq j)$~~ & ~~$|\psi\rangle_{kl}$ ($k\neq l\neq i\neq j$)~~ & ~~ Basic constraints  ~~  &  ~~ Extended constraints ~~\\
  \hline
 ~~$Z_1=i,Z_2=-i$~~  & ~~$|\psi\rangle_{34}=\frac{1}{\sqrt{2}}(i|00\rangle-i|22\rangle)$~~ &  ~~$X_3^2X_4^2=-1$~~ & ~~$X_3^2X_4^2=-1$~~\\
 ~~$Z_1=i,Z_3=-1$~~  & ~~$|\psi\rangle_{24}=\frac{1}{2}(i|01\rangle-i|10\rangle+i|23\rangle-i|32\rangle)$~~ &  ~~$X_2X_4^3=-1$~~ & ~~$X_2^2X_4^2=1$~~ \\
 ~~$Z_1=i,Z_4=-1$~~  & ~~$|\psi\rangle_{23}=\frac{1}{2}(i|01\rangle+i|10\rangle-i|23\rangle-i|32\rangle)$~~ &  ~~$X_2^2X_3^2=-1$~~ & ~~$X_2^2X_3^2=-1$~~ \\
 ~~$Z_2=-i,Z_3=-1$~~  & ~~$|\psi\rangle_{14}=\frac{1}{2}(i|03\rangle-i|12\rangle+i|21\rangle-i|30\rangle)$~~ &  ~~$X_1X_4^3=-1$~~ &  ~~$X_1^2X_4^2=1$~~\\
 ~~$Z_2=-i,Z_4=-1$~~  & ~~$|\psi\rangle_{13}=\frac{1}{2}(-i|03\rangle-i|12\rangle+i|21\rangle+i|30\rangle)$~~ &  ~~$X_1^2X_3^2=-1$~~ & ~~$X_1^2X_3^2=-1$~~ \\
 ~~$Z_3=-1,Z_4=-1$~~  & ~~$|\psi\rangle_{12}=\frac{1}{2}(-|00\rangle-i|13\rangle+|22\rangle+i|31\rangle)$~~ &  ~~$X_1X_2^3=-i$~~ & ~~$X_1^2X_2^2=-1$~~ \\
     \hline
 ~~$Z_1=-1,Z_2=i$~~  & ~~$|\psi\rangle_{34}=\frac{1}{2}(-i|01\rangle+i|10\rangle-i|23\rangle-i|32\rangle)$~~ &  ~~$\text{---}$~~ & ~~$\text{---}$~~\\
 ~~$Z_1=-1,Z_3=-i$~~  & ~~$|\psi\rangle_{24}=\frac{1}{2}(-i|03\rangle-i|12\rangle+i|21\rangle+i|30\rangle)$~~ &  ~~$X_2^2X_4^2=-1$~~ & ~~$X_2^2X_4^2=-1$~~ \\
 ~~$Z_1=-1,Z_4=-1$~~  & ~~$|\psi\rangle_{23}=\frac{1}{2}(-|00\rangle-i|13\rangle+|22\rangle+i|31\rangle)$~~ &  ~~$X_2X_3^3=-i$~~ & ~~$X_2^2X_3^2=-1$~~ \\
 ~~$Z_2=i,Z_3=-i$~~  & ~~$|\psi\rangle_{14}=\frac{1}{\sqrt{2}}(i|00\rangle-i|22\rangle)$~~ &  ~~$X_1^2X_4^2=-1$~~ &  ~~$X_1^2X_4^2=-1$~~\\
 ~~$Z_2=i,Z_4=-1$~~  & ~~$|\psi\rangle_{13}=\frac{1}{2}(-i|01\rangle+i|10\rangle-i|23\rangle+i|32\rangle)$~~ &  ~~$X_1X_3^3=-1$~~ & ~~$X_1^2X_3^2=1$~~ \\
 ~~$Z_3=-i,Z_4=-1$~~  & ~~$|\psi\rangle_{12}=\frac{1}{2}(-i|03\rangle-i|12\rangle-i|21\rangle+i|30\rangle)$~~ &  ~~$\text{---}$~~ & ~~$\text{---}$~~ \\
     \hline
 ~~$Z_1=-1,Z_2=-1$~~  & ~~$|\psi\rangle_{34}=\frac{1}{2}(-|00\rangle-i|13\rangle+|22\rangle+i|31\rangle)$~~ &  ~~$X_3X_4^3=-i$~~ & ~~$X_3^2X_4^2=-1$~~\\
 ~~$Z_1=-1,Z_3=i$~~  & ~~$|\psi\rangle_{24}=\frac{1}{2}(-i|01\rangle+i|10\rangle-i|23\rangle+i|32\rangle)$~~ &  ~~$X_2X_4^3=-1$~~ & ~~$X_2^2X_4^2=1$~~ \\
 ~~$Z_1=-1,Z_4=-i$~~  & ~~$|\psi\rangle_{23}=\frac{1}{2}(-i|03\rangle-i|12\rangle-i|21\rangle-i|30\rangle)$~~ &  ~~$X_2X_3^3=1$~~ & ~~$X_2^2X_3^2=1$~~ \\
 ~~$Z_2=-1,Z_3=i$~~  & ~~$|\psi\rangle_{14}=\frac{1}{2}(i|01\rangle-i|10\rangle-i|23\rangle+i|32\rangle)$~~ &  ~~$X_1^2X_4^2=-1$~~ &  ~~$X_1^2X_4^2=-1$~~\\
 ~~$Z_2=-1,Z_4=-i$~~  & ~~$|\psi\rangle_{13}=\frac{1}{2}(i|03\rangle+i|12\rangle-i|21\rangle-i|30\rangle)$~~ &  ~~$X_1^2X_3^2=-1$~~ & ~~$X_1^2X_3^2=-1$~~ \\
 ~~$Z_3=i,Z_4=-i$~~  & ~~$|\psi\rangle_{12}=\frac{1}{\sqrt{2}}(i|00\rangle-i|22\rangle)$~~ &  ~~$X_1^2X_2^2=-1$~~ & ~~$X_1^2X_2^2=-1$~~ \\
     \hline
 ~~$Z_1=-i,Z_2=-1$~~  & ~~$|\psi\rangle_{34}=\frac{1}{2}(-i|03\rangle+i|12\rangle-i|21\rangle-i|30\rangle)$~~ &  ~~$\text{---}$~~ & ~~$\text{---}$~~\\
 ~~$Z_1=-i,Z_3=-1$~~  & ~~$|\psi\rangle_{24}=\frac{1}{2}(i|03\rangle+i|12\rangle-i|21\rangle-i|30\rangle)$~~ &  ~~$X_2^2X_4^2=-1$~~ & ~~$X_2^2X_4^2=-1$~~ \\
 ~~$Z_1=-i,Z_4=i$~~  & ~~$|\psi\rangle_{23}=\frac{1}{\sqrt{2}}(i|00\rangle-i|22\rangle)$~~ &  ~~$X_2^2X_3^2=-1$~~ & ~~$X_2^2X_3^2=-1$~~ \\
 ~~$Z_2=-1,Z_3=-1$~~  & ~~$|\psi\rangle_{14}=\frac{1}{2}(-|00\rangle+i|13\rangle+|22\rangle-i|31\rangle)$~~ &  ~~$X_1X_4^3=i$~~ &  ~~$X_1^2X_4^2=-1$~~\\
 ~~$Z_2=-1,Z_4=i$~~  & ~~$|\psi\rangle_{13}=\frac{1}{2}(i|01\rangle-i|10\rangle+i|23\rangle-i|32\rangle)$~~ &  ~~$X_1X_3^3=-1$~~ & ~~$X_1^2X_3^2=1$~~ \\
 ~~$Z_3=-1,Z_4=i$~~  & ~~$|\psi\rangle_{12}=\frac{1}{2}(i|01\rangle+i|10\rangle+i|23\rangle-i|32\rangle)$~~ & ~~$\text{---}$~~ & ~~$\text{---}$~~\\
     \hline
 ~~$Z_1=i,Z_2=-1$~~  & ~~$|\psi\rangle_{34}=\frac{1}{2}(-i|01\rangle-i|10\rangle+i|23\rangle-i|32\rangle)$~~ &  ~~$\text{---}$~~ & ~~$\text{---}$~~\\
 ~~$Z_1=i,Z_3=-i$~~  & ~~$|\psi\rangle_{24}=\frac{1}{\sqrt{2}}(i|00\rangle-i|22\rangle)$~~ &  ~~$X_2^2X_4^2=-1$~~ & ~~$X_2^2X_4^2=-1$~~ \\
 ~~$Z_1=i,Z_4=-1$~~  & ~~$|\psi\rangle_{23}=\frac{1}{2}(i|01\rangle+i|10\rangle-i|23\rangle-i|32\rangle)$~~ &  ~~$X_2^2X_3^2=-1$~~ & ~~$X_2^2X_3^2=-1$~~ \\
 ~~$Z_2=-1,Z_3=-i$~~  & ~~$|\psi\rangle_{14}=\frac{1}{2}(i|03\rangle-i|12\rangle+i|21\rangle-i|30\rangle)$~~ &  ~~$X_1X_4^3=-1$~~ &  ~~$X_1^2X_4^2=1$~~\\
 ~~$Z_2=-1,Z_4=-1$~~  & ~~$|\psi\rangle_{13}=\frac{1}{2}(-|00\rangle-i|13\rangle+|22\rangle+i|31\rangle)$~~ &  ~~$X_1X_3^3=-i$~~ & ~~$X_1^2X_3^2=-1$~~ \\
 ~~$Z_3=-i,Z_4=-1$~~  & ~~$|\psi\rangle_{12}=\frac{1}{2}(-i|03\rangle-i|12\rangle-i|21\rangle+i|30\rangle)$~~ &  ~~$\text{---}$~~ & ~~$\text{---}$~~ \\
     \hline
 ~~$Z_1=-1,Z_2=i$~~  & ~~$|\psi\rangle_{34}=\frac{1}{2}(-i|01\rangle+i|10\rangle-i|23\rangle-i|32\rangle)$~~ & ~~$\text{---}$~~ & ~~$\text{---}$~~\\
 ~~$Z_1=-1,Z_3=-1$~~  & ~~$|\psi\rangle_{24}=\frac{1}{2}(-|00\rangle-i|13\rangle+|22\rangle+i|31\rangle)$~~ &  ~~$X_2X_4^3=-i$~~ & ~~$X_2^2X_4^2=-1$~~ \\
 ~~$Z_1=-1,Z_4=-i$~~  & ~~$|\psi\rangle_{23}=\frac{1}{2}(-i|03\rangle-i|12\rangle-i|21\rangle-i|30\rangle)$~~ &  ~~$X_2X_3^3=1$~~ & ~~$X_2^2X_3^2=1$~~ \\
 ~~$Z_2=i,Z_3=-1$~~  & ~~$|\psi\rangle_{14}=\frac{1}{2}(i|01\rangle-i|10\rangle-i|23\rangle+i|32\rangle)$~~ &  ~~$X_1^2X_4^2=-1$~~ &  ~~$X_1^2X_4^2=-1$~~\\
 ~~$Z_2=i,Z_4=-i$~~  & ~~$|\psi\rangle_{13}=\frac{1}{\sqrt{2}}(i|00\rangle-i|22\rangle)$~~ &  ~~$X_1^2X_3^2=-1$~~ & ~~$X_1^2X_3^2=-1$~~ \\
 ~~$Z_3=-1,Z_4=-i$~~  & ~~$|\psi\rangle_{12}=\frac{1}{2}(i|03\rangle+i|12\rangle-i|21\rangle+i|30\rangle)$~~ &  ~~$\text{---}$~~ & ~~$\text{---}$~~ \\
     \hline
 \hline
    \end{tabular}
\end{table*}

\begin{table*}\caption{Six groups of Hardy-like conditions for type VI-B, in which the measurement results for $Z_1$, $Z_2$, $Z_3$, $Z_4$ arise from the contributions of the terms $i|3122\rangle$, $i|2312\rangle$, $i|2231\rangle$, $i|1223\rangle$, $i|3212\rangle$, $i|2321\rangle$, respectively.} \label{TB10}
    \centering
    \centering
    \begin{tabular}{c|ccc}
    \hline
    \hline
        ~~$Z_i,Z_j$ $(i\neq j)$~~ & ~~$|\psi\rangle_{kl}$ ($k\neq l\neq i\neq j$)~~ & ~~ Basic constraints  ~~  &  ~~ Extended constraints ~~\\
  \hline
 ~~$Z_1=-i,Z_2=i$~~  & ~~$|\psi\rangle_{34}=\frac{1}{\sqrt{2}}(-i|00\rangle+i|22\rangle)$~~ &  ~~$X_3^2X_4^2=-1$~~ & ~~$X_3^2X_4^2=-1$~~\\
 ~~$Z_1=-i,Z_3=-1$~~  & ~~$|\psi\rangle_{24}=\frac{1}{2}(i|03\rangle+i|12\rangle-i|21\rangle-i|30\rangle)$~~ &  ~~$X_2^2X_4^2=-1$~~ & ~~$X_2^2X_4^2=-1$~~ \\
 ~~$Z_1=-i,Z_4=-1$~~  & ~~$|\psi\rangle_{23}=\frac{1}{2}(i|03\rangle+i|12\rangle+i|21\rangle+i|30\rangle)$~~ &  ~~$X_2X_3^3=1$~~ & ~~$X_2^2X_3^2=1$~~ \\
 ~~$Z_2=i,Z_3=-1$~~  & ~~$|\psi\rangle_{14}=\frac{1}{2}(i|01\rangle-i|10\rangle-i|23\rangle+i|32\rangle)$~~ &  ~~$X_1^2X_4^2=-1$~~ &  ~~$X_1^2X_4^2=-1$~~\\
 ~~$Z_2=i,Z_4=-1$~~  & ~~$|\psi\rangle_{13}=\frac{1}{2}(-i|01\rangle+i|10\rangle-i|23\rangle+i|32\rangle)$~~ &  ~~$X_1X_3^3=-1$~~ & ~~$X_1^2X_3^2=1$~~ \\
 ~~$Z_3=-1,Z_4=-1$~~  & ~~$|\psi\rangle_{12}=\frac{1}{2}(-|00\rangle-i|13\rangle+|22\rangle+i|31\rangle)$~~ &  ~~$X_1X_2^3=-i$~~ & ~~$X_1^2X_2^2=-1$~~ \\
     \hline
 ~~$Z_1=-1,Z_2=-i$~~  & ~~$|\psi\rangle_{34}=\frac{1}{2}(-i|03\rangle+i|12\rangle+i|21\rangle+i|30\rangle)$~~ &  ~~$\text{---}$~~ & ~~$\text{---}$~~\\
 ~~$Z_1=-1,Z_3=i$~~  & ~~$|\psi\rangle_{24}=\frac{1}{2}(-i|01\rangle+i|10\rangle-i|23\rangle+i|32\rangle)$~~ &  ~~$X_2X_4^3=-1$~~ & ~~$X_2^2X_4^2=1$~~ \\
 ~~$Z_1=-1,Z_4=-1$~~  & ~~$|\psi\rangle_{23}=\frac{1}{2}(-|00\rangle-i|13\rangle+|22\rangle+i|31\rangle)$~~ &  ~~$X_2X_3^3=-i$~~ & ~~$X_2^2X_3^2=-1$~~ \\
 ~~$Z_2=-i,Z_3=i$~~  & ~~$|\psi\rangle_{14}=\frac{1}{\sqrt{2}}(-i|00\rangle+i|22\rangle)$~~ &  ~~$X_1^2X_4^2=-1$~~ &  ~~$X_1^2X_4^2=-1$~~\\
 ~~$Z_2=-i,Z_4=-1$~~  & ~~$|\psi\rangle_{13}=\frac{1}{2}(-i|03\rangle-i|12\rangle+i|21\rangle+i|30\rangle)$~~ &  ~~$X_1^2X_3^2=-1$~~ & ~~$X_1^2X_3^2=-1$~~ \\
 ~~$Z_3=i,Z_4=-1$~~  & ~~$|\psi\rangle_{12}=\frac{1}{2}(-i|01\rangle+i|10\rangle+i|23\rangle+i|32\rangle)$~~ &  ~~$\text{---}$~~ & ~~$\text{---}$~~ \\
     \hline
 ~~$Z_1=-1,Z_2=-1$~~  & ~~$|\psi\rangle_{34}=\frac{1}{2}(-|00\rangle-i|13\rangle+|22\rangle+i|31\rangle)$~~ &  ~~$X_3X_4^3=-i$~~ & ~~$X_3^2X_4^2=-1$~~\\
 ~~$Z_1=-1,Z_3=-i$~~  & ~~$|\psi\rangle_{24}=\frac{1}{2}(-i|03\rangle-i|12\rangle+i|21\rangle+i|30\rangle)$~~ &  ~~$X_2^2X_4^2=-1$~~ & ~~$X_2^2X_4^2=-1$~~ \\
 ~~$Z_1=-1,Z_4=i$~~  & ~~$|\psi\rangle_{23}=\frac{1}{2}(-i|01\rangle-i|10\rangle+i|23\rangle+i|32\rangle)$~~ &  ~~$X_2^2X_3^2=-1$~~ & ~~$X_2^2X_3^2=-1$~~ \\
 ~~$Z_2=-1,Z_3=-i$~~  & ~~$|\psi\rangle_{14}=\frac{1}{2}(i|03\rangle-i|12\rangle+i|21\rangle-i|30\rangle)$~~ &  ~~$X_1X_4^3=-1$~~ &  ~~$X_1^2X_4^2=1$~~\\
 ~~$Z_2=-1,Z_4=i$~~  & ~~$|\psi\rangle_{13}=\frac{1}{2}(i|01\rangle-i|10\rangle+i|23\rangle-i|32\rangle)$~~ &  ~~$X_1X_3^3=-1$~~ & ~~$X_1^2X_3^2=1$~~ \\
 ~~$Z_3=-i,Z_4=i$~~  & ~~$|\psi\rangle_{12}=\frac{1}{\sqrt{2}}(-i|00\rangle+i|22\rangle)$~~ &  ~~$X_1^2X_2^2=-1$~~ & ~~$X_1^2X_2^2=-1$~~ \\
     \hline
 ~~$Z_1=i,Z_2=-1$~~  & ~~$|\psi\rangle_{34}=\frac{1}{2}(-i|01\rangle-i|10\rangle+i|23\rangle-i|32\rangle)$~~ &  ~~$\text{---}$~~ & ~~$\text{---}$~~\\
 ~~$Z_1=i,Z_3=-1$~~  & ~~$|\psi\rangle_{24}=\frac{1}{2}(i|01\rangle-i|10\rangle+i|23\rangle-i|32\rangle)$~~ &  ~~$X_2X_4^3=-1$~~ & ~~$X_2^2X_4^2=1$~~ \\
 ~~$Z_1=i,Z_4=-i$~~  & ~~$|\psi\rangle_{23}=\frac{1}{\sqrt{2}}(-i|00\rangle+i|22\rangle)$~~ &  ~~$X_2^2X_3^2=-1$~~ & ~~$X_2^2X_3^2=-1$~~ \\
 ~~$Z_2=-1,Z_3=-1$~~  & ~~$|\psi\rangle_{14}=\frac{1}{2}(-|00\rangle+i|13\rangle+|22\rangle-i|31\rangle)$~~ &  ~~$X_1X_4^3=i$~~ &  ~~$X_1^2X_4^2=-1$~~\\
 ~~$Z_2=-1,Z_4=-i$~~  & ~~$|\psi\rangle_{13}=\frac{1}{2}(i|03\rangle+i|12\rangle-i|21\rangle-i|30\rangle)$~~ &  ~~$X_1^2X_3^2=-1$~~ & ~~$X_1^2X_3^2=-1$~~ \\
 ~~$Z_3=-1,Z_4=-i$~~  & ~~$|\psi\rangle_{12}=\frac{1}{2}(i|03\rangle+i|12\rangle-i|21\rangle+i|30\rangle)$~~ & ~~$\text{---}$~~ & ~~$\text{---}$~~\\
     \hline
 ~~$Z_1=-i,Z_2=-1$~~  & ~~$|\psi\rangle_{34}=\frac{1}{2}(-i|03\rangle+i|12\rangle-i|21\rangle-i|30\rangle)$~~ &  ~~$\text{---}$~~ & ~~$\text{---}$~~\\
 ~~$Z_1=-i,Z_3=i$~~  & ~~$|\psi\rangle_{24}=\frac{1}{\sqrt{2}}(-i|00\rangle+i|22\rangle)$~~ &  ~~$X_2^2X_4^2=-1$~~ & ~~$X_2^2X_4^2=-1$~~ \\
 ~~$Z_1=-i,Z_4=-1$~~  & ~~$|\psi\rangle_{23}=\frac{1}{2}(i|03\rangle+i|12\rangle+i|21\rangle+i|30\rangle)$~~ &  ~~$X_2X_3^3=1$~~ & ~~$X_2^2X_3^2=1$~~ \\
 ~~$Z_2=-1,Z_3=i$~~  & ~~$|\psi\rangle_{14}=\frac{1}{2}(i|01\rangle-i|10\rangle-i|23\rangle+i|32\rangle)$~~ &  ~~$X_1^2X_4^2=-1$~~ &  ~~$X_1^2X_4^2=-1$~~\\
 ~~$Z_2=-1,Z_4=-1$~~  & ~~$|\psi\rangle_{13}=\frac{1}{2}(-|00\rangle-i|13\rangle+|22\rangle+i|31\rangle)$~~ &  ~~$X_1X_3^3=-i$~~ & ~~$X_1^2X_3^2=-1$~~ \\
 ~~$Z_3=i,Z_4=-1$~~  & ~~$|\psi\rangle_{12}=\frac{1}{2}(-i|01\rangle+i|10\rangle+i|23\rangle+i|32\rangle)$~~ &  ~~$\text{---}$~~ & ~~$\text{---}$~~ \\
     \hline
 ~~$Z_1=-1,Z_2=-i$~~  & ~~$|\psi\rangle_{34}=\frac{1}{2}(-i|03\rangle+i|12\rangle+i|21\rangle+i|30\rangle)$~~ & ~~$\text{---}$~~ & ~~$\text{---}$~~\\
 ~~$Z_1=-1,Z_3=-1$~~  & ~~$|\psi\rangle_{24}=\frac{1}{2}(-|00\rangle-i|13\rangle+|22\rangle+i|31\rangle)$~~ &  ~~$X_2X_4^3=-i$~~ & ~~$X_2^2X_4^2=-1$~~ \\
 ~~$Z_1=-1,Z_4=i$~~  & ~~$|\psi\rangle_{23}=\frac{1}{2}(-i|01\rangle-i|10\rangle+i|23\rangle+i|32\rangle)$~~ &  ~~$X_2^2X_3^2=-1$~~ & ~~$X_2^2X_3^2=-1$~~ \\
 ~~$Z_2=-i,Z_3=-1$~~  & ~~$|\psi\rangle_{14}=\frac{1}{2}(i|03\rangle-i|12\rangle+i|21\rangle-i|30\rangle)$~~ &  ~~$X_1X_4^3=-1$~~ &  ~~$X_1^2X_4^2=1$~~\\
 ~~$Z_2=-i,Z_4=i$~~  & ~~$|\psi\rangle_{13}=\frac{1}{\sqrt{2}}(-i|00\rangle+i|22\rangle)$~~ &  ~~$X_1^2X_3^2=-1$~~ & ~~$X_1^2X_3^2=-1$~~ \\
 ~~$Z_3=-1,Z_4=i$~~  & ~~$|\psi\rangle_{12}=\frac{1}{2}(i|01\rangle+i|10\rangle+i|23\rangle-i|32\rangle)$~~ &  ~~$\text{---}$~~ & ~~$\text{---}$~~ \\
     \hline
 \hline
    \end{tabular}
\end{table*}


\begin{thebibliography}{13}%
\makeatletter
\providecommand \@ifxundefined [1]{%
 \@ifx{#1\undefined}
}%
\providecommand \@ifnum [1]{%
 \ifnum #1\expandafter \@firstoftwo
 \else \expandafter \@secondoftwo
 \fi
}%
\providecommand \@ifx [1]{%
 \ifx #1\expandafter \@firstoftwo
 \else \expandafter \@secondoftwo
 \fi
}%
\providecommand \natexlab [1]{#1}%
\providecommand \enquote  [1]{``#1''}%
\providecommand \bibnamefont  [1]{#1}%
\providecommand \bibfnamefont [1]{#1}%
\providecommand \citenamefont [1]{#1}%
\providecommand \href@noop [0]{\@secondoftwo}%
\providecommand \href [0]{\begingroup \@sanitize@url \@href}%
\providecommand \@href[1]{\@@startlink{#1}\@@href}%
\providecommand \@@href[1]{\endgroup#1\@@endlink}%
\providecommand \@sanitize@url [0]{\catcode `\\12\catcode `\$12\catcode
  `\&12\catcode `\#12\catcode `\^12\catcode `\_12\catcode `\%12\relax}%
\providecommand \@@startlink[1]{}%
\providecommand \@@endlink[0]{}%
\providecommand \url  [0]{\begingroup\@sanitize@url \@url }%
\providecommand \@url [1]{\endgroup\@href {#1}{\urlprefix }}%
\providecommand \urlprefix  [0]{URL }%
\providecommand \Eprint [0]{\href }%
\providecommand \doibase [0]{http://dx.doi.org/}%
\providecommand \selectlanguage [0]{\@gobble}%
\providecommand \bibinfo  [0]{\@secondoftwo}%
\providecommand \bibfield  [0]{\@secondoftwo}%
\providecommand \translation [1]{[#1]}%
\providecommand \BibitemOpen [0]{}%
\providecommand \bibitemStop [0]{}%
\providecommand \bibitemNoStop [0]{.\EOS\space}%
\providecommand \EOS [0]{\spacefactor3000\relax}%
\providecommand \BibitemShut  [1]{\csname bibitem#1\endcsname}%
\let\auto@bib@innerbib\@empty
\bibitem [{\citenamefont {Bell}(1964)}]{Bell}%
  \BibitemOpen
  \bibfield  {author} {\bibinfo {author} {\bibfnamefont {J.~S.}\ \bibnamefont
  {Bell}},\ }\bibfield  {title} {\enquote {\bibinfo {title} {On the {Einstein
  Podolsky Rosen paradox}},}\ }\href {\doibase
  10.1103/PhysicsPhysiqueFizika.1.195} {\bibfield  {journal} {\bibinfo
  {journal} {Physics Physique Fizika}\ }\textbf {\bibinfo {volume} {1}},\
  \bibinfo {pages} {195--200} (\bibinfo {year} {1964})}\BibitemShut {NoStop}%
\bibitem [{\citenamefont {Bell}(1966)}]{Bell2}%
  \BibitemOpen
  \bibfield  {author} {\bibinfo {author} {\bibfnamefont {John~S.}\ \bibnamefont
  {Bell}},\ }\bibfield  {title} {\enquote {\bibinfo {title} {On the problem of
  hidden variables in quantum mechanics},}\ }\href {\doibase
  10.1103/RevModPhys.38.447} {\bibfield  {journal} {\bibinfo  {journal} {Rev.
  Mod. Phys.}\ }\textbf {\bibinfo {volume} {38}},\ \bibinfo {pages} {447--452}
  (\bibinfo {year} {1966})}\BibitemShut {NoStop}%
\bibitem [{\citenamefont {Clauser}\ \emph {et~al.}(1969)\citenamefont
  {Clauser}, \citenamefont {Horne}, \citenamefont {Shimony},\ and\
  \citenamefont {Holt}}]{CHSH}%
  \BibitemOpen
  \bibfield  {author} {\bibinfo {author} {\bibfnamefont {John~F.}\ \bibnamefont
  {Clauser}}, \bibinfo {author} {\bibfnamefont {Michael~A.}\ \bibnamefont
  {Horne}}, \bibinfo {author} {\bibfnamefont {Abner}\ \bibnamefont {Shimony}},
  \ and\ \bibinfo {author} {\bibfnamefont {Richard~A.}\ \bibnamefont {Holt}},\
  }\bibfield  {title} {\enquote {\bibinfo {title} {Proposed experiment to test
  local hidden-variable theories},}\ }\href {\doibase
  10.1103/PhysRevLett.23.880} {\bibfield  {journal} {\bibinfo  {journal} {Phys.
  Rev. Lett.}\ }\textbf {\bibinfo {volume} {23}},\ \bibinfo {pages} {880--884}
  (\bibinfo {year} {1969})}\BibitemShut {NoStop}%
\bibitem [{\citenamefont {Brunner}\ \emph {et~al.}(2014)\citenamefont
  {Brunner}, \citenamefont {Cavalcanti}, \citenamefont {Pironio}, \citenamefont
  {Scarani},\ and\ \citenamefont {Wehner}}]{Bell-Nonlocality-RMP-2014}%
  \BibitemOpen
  \bibfield  {author} {\bibinfo {author} {\bibfnamefont {Nicolas}\ \bibnamefont
  {Brunner}}, \bibinfo {author} {\bibfnamefont {Daniel}\ \bibnamefont
  {Cavalcanti}}, \bibinfo {author} {\bibfnamefont {Stefano}\ \bibnamefont
  {Pironio}}, \bibinfo {author} {\bibfnamefont {Valerio}\ \bibnamefont
  {Scarani}}, \ and\ \bibinfo {author} {\bibfnamefont {Stephanie}\ \bibnamefont
  {Wehner}},\ }\bibfield  {title} {\enquote {\bibinfo {title} {Bell
  nonlocality},}\ }\href {\doibase 10.1103/RevModPhys.86.419} {\bibfield
  {journal} {\bibinfo  {journal} {Rev. Mod. Phys.}\ }\textbf {\bibinfo {volume}
  {86}},\ \bibinfo {pages} {419--478} (\bibinfo {year} {2014})}\BibitemShut
  {NoStop}%
\bibitem [{\citenamefont {Mermin}(1990)}]{all-vs-nothing}%
  \BibitemOpen
  \bibfield  {author} {\bibinfo {author} {\bibfnamefont {N.~D.}\ \bibnamefont
  {Mermin}},\ }\bibfield  {title} {\enquote {\bibinfo {title} {Extreme quantum
  entanglement in a superposition of macroscopically distinct states},}\ }\href
  {\doibase 10.1103/PhysRevLett.65.1838} {\bibfield  {journal} {\bibinfo
  {journal} {Phys. Rev. Lett.}\ }\textbf {\bibinfo {volume} {65}},\ \bibinfo
  {pages} {1838--1840} (\bibinfo {year} {1990})}\BibitemShut {NoStop}%
\bibitem [{\citenamefont {Greenberger}\ \emph {et~al.}(1989)\citenamefont
  {Greenberger}, \citenamefont {Horne},\ and\ \citenamefont
  {Zeilinger}}]{GHZ1989}%
  \BibitemOpen
  \bibfield  {author} {\bibinfo {author} {\bibfnamefont {Daniel~M}\
  \bibnamefont {Greenberger}}, \bibinfo {author} {\bibfnamefont {Michael~A}\
  \bibnamefont {Horne}}, \ and\ \bibinfo {author} {\bibfnamefont {Anton}\
  \bibnamefont {Zeilinger}},\ }\bibfield  {title} {\enquote {\bibinfo {title}
  {Going beyond {Bell's} theorem},}\ }in\ \href {\doibase
  10.1007/978-94-017-0849-4} {\emph {\bibinfo {booktitle} {Bell's theorem,
  quantum theory and conceptions of the universe}}}\ (\bibinfo  {publisher}
  {Springer},\ \bibinfo {year} {1989})\ pp.\ \bibinfo {pages}
  {69--72}\BibitemShut {NoStop}%
\bibitem [{\citenamefont {Greenberger}\ \emph {et~al.}(1990)\citenamefont
  {Greenberger}, \citenamefont {Horne}, \citenamefont {Shimony},\ and\
  \citenamefont {Zeilinger}}]{GHZ1990}%
  \BibitemOpen
  \bibfield  {author} {\bibinfo {author} {\bibfnamefont {Daniel~M.}\
  \bibnamefont {Greenberger}}, \bibinfo {author} {\bibfnamefont {Michael~A.}\
  \bibnamefont {Horne}}, \bibinfo {author} {\bibfnamefont {Abner}\ \bibnamefont
  {Shimony}}, \ and\ \bibinfo {author} {\bibfnamefont {Anton}\ \bibnamefont
  {Zeilinger}},\ }\bibfield  {title} {\enquote {\bibinfo {title} {Bell's
  theorem without inequalities},}\ }\href {\doibase 10.1119/1.16243} {\bibfield
   {journal} {\bibinfo  {journal} {American Journal of Physics}\ }\textbf
  {\bibinfo {volume} {58}},\ \bibinfo {pages} {1131--1143} (\bibinfo {year}
  {1990})}\BibitemShut {NoStop}%
\bibitem [{\citenamefont {Cabello}(2001)}]{Cabello2001}%
  \BibitemOpen
  \bibfield  {author} {\bibinfo {author} {\bibfnamefont {Ad\'an}\ \bibnamefont
  {Cabello}},\ }\bibfield  {title} {\enquote {\bibinfo {title} {Bell's theorem
  without inequalities and without probabilities for two observers},}\ }\href
  {\doibase 10.1103/PhysRevLett.86.1911} {\bibfield  {journal} {\bibinfo
  {journal} {Phys. Rev. Lett.}\ }\textbf {\bibinfo {volume} {86}},\ \bibinfo
  {pages} {1911--1914} (\bibinfo {year} {2001})}\BibitemShut {NoStop}%
\bibitem [{\citenamefont {Hardy}(1992)}]{Hardy92}%
  \BibitemOpen
  \bibfield  {author} {\bibinfo {author} {\bibfnamefont {Lucien}\ \bibnamefont
  {Hardy}},\ }\bibfield  {title} {\enquote {\bibinfo {title} {Quantum
  mechanics, local realistic theories, and lorentz-invariant realistic
  theories},}\ }\href {\doibase 10.1103/PhysRevLett.68.2981} {\bibfield
  {journal} {\bibinfo  {journal} {Phys. Rev. Lett.}\ }\textbf {\bibinfo
  {volume} {68}},\ \bibinfo {pages} {2981--2984} (\bibinfo {year}
  {1992})}\BibitemShut {NoStop}%
\bibitem [{\citenamefont {Hardy}(1993)}]{Hardy93}%
  \BibitemOpen
  \bibfield  {author} {\bibinfo {author} {\bibfnamefont {Lucien}\ \bibnamefont
  {Hardy}},\ }\bibfield  {title} {\enquote {\bibinfo {title} {Nonlocality for
  two particles without inequalities for almost all entangled states},}\ }\href
  {\doibase 10.1103/PhysRevLett.71.1665} {\bibfield  {journal} {\bibinfo
  {journal} {Phys. Rev. Lett.}\ }\textbf {\bibinfo {volume} {71}},\ \bibinfo
  {pages} {1665--1668} (\bibinfo {year} {1993})}\BibitemShut {NoStop}%
\bibitem [{\citenamefont {Tang}(2022{\natexlab{a}})}]{Tang-DAVN-2022}%
  \BibitemOpen
  \bibfield  {author} {\bibinfo {author} {\bibfnamefont {Weidong}\ \bibnamefont
  {Tang}},\ }\bibfield  {title} {\enquote {\bibinfo {title} {Deterministic
  all-versus-nothing proofs of Bell nonlocality based on non-stabilizer
  states},}\ }\href {https://arxiv.org/abs/2201.01886} {\bibfield  {journal}
  {\bibinfo  {journal} {preprint arXiv:2201.01886}\ } (\bibinfo {year}
  {2022}{\natexlab{a}})}\BibitemShut {NoStop}%
\bibitem [{Her()}]{Hermitian-measurement}%
  \BibitemOpen
  \href@noop {} {}\bibinfo {note} {In fact, each $Z_j$ is just a unitary rather
  than Hermitian operator. However, it can be represented by $Z_j=e^{iH_j}$,
  where $H_j$ is a Hermitian operator, indicating that the measurement of $Z_j$
  can be equivalently to the measurement of $H_j$, and for simplicity thus here
  we still use the terminology ``measurement of $Z_j$".}\BibitemShut {Stop}%
\bibitem [{\citenamefont {Tang}(2022{\natexlab{b}})}]{Tang2022}%
  \BibitemOpen
  \bibfield  {author} {\bibinfo {author} {\bibfnamefont {Weidong}\ \bibnamefont
  {Tang}},\ }\bibfield  {title} {\enquote {\bibinfo {title} {Hardy-like quantum
  pigeonhole paradox and the projected-coloring graph state},}\ }\href
  {\doibase 10.1103/PhysRevA.105.032457} {\bibfield  {journal} {\bibinfo
  {journal} {Phys. Rev. A}\ }\textbf {\bibinfo {volume} {105}},\ \bibinfo
  {pages} {032457} (\bibinfo {year} {2022}{\natexlab{b}})}\BibitemShut
  {NoStop}%
\end{thebibliography}
\end{document}